\documentclass[useAMS,usenatbib]{mn2e}
\input psfig.sty

\def \aj {AJ}
\def \apj {ApJ}
\def \apjl {ApJL}
\def \mnras {MNRAS}
\def \apjs {ApJS}
\def \aap {A\&A}
\def \nat {Nature}

\def \pasp {PASP}
\def \na {NewA}
\def \etal {et~al.~}

\def \spose#1{\hbox  to 0pt{#1\hss}}  
\def \lta{\mathrel{\spose{\lower 3pt\hbox{$\sim$}}\raise  2.0pt\hbox{$<$}}}
\def \gta{\mathrel{\spose{\lower  3pt\hbox{$\sim$}}\raise 2.0pt\hbox{$>$}}}

\def \kms {\ifmmode  \,\rm km\,s^{-1} \else $\,\rm km\,s^{-1}  $ \fi }
\def \kpc {\ifmmode  {\rm kpc}  \else ${\rm  kpc}$ \fi  }  
\def \hkpc {\ifmmode  {h^{-1}\rm kpc}  \else ${h^{-1}\rm kpc}$ \fi  }  
\def \hMpc {\ifmmode  {h^{-1}\rm Mpc}  \else ${h^{-1}\rm Mpc}$ \fi  }  
\def \Msun {\ifmmode {\rm M}_{\odot} \else ${\rm M}_{\odot}$ \fi} 
\def \hMsun {\ifmmode h^{-1}\,\rm M_{\odot} \else $h^{-1}\,\rm M_{\odot}$ \fi}
\def \hhMsun {\ifmmode h^{-2}\,\rm M_{\odot}\else $h^{-2}\,\rm M_{\odot}$ \fi}
\def \Lsun {\ifmmode L_{\odot} \else $L_{\odot}$ \fi} 
\def \hhLsun {\ifmmode h^{-2}\,\rm L_{\odot} \else $h^{-2}\,\rm L_{\odot}$ \fi}
\def \Mstar {\ifmmode M_{\rm star} \else $M_{\rm star}$ \fi} 
\def \Mgas {\ifmmode M_{\rm gas} \else $M_{\rm gas}$ \fi} 

\def \LCDM {\ifmmode \Lambda{\rm CDM} \else $\Lambda{\rm CDM}$ \fi}
\def \sig8 {\ifmmode \sigma_8 \else $\sigma_8$ \fi} 
\def \Omegam {\ifmmode \Omega_{\rm m} \else $\Omega_{\rm m}$ \fi} 
\def \Omegab {\ifmmode \Omega_{\rm b} \else $\Omega_{\rm b}$ \fi} 
\def \OmegaL {\ifmmode \Omega_{\rm \Lambda} \else $\Omega_{\rm \Lambda}$\fi} 
\def \Deltavir {\ifmmode \Delta_{\rm vir} \else $\Delta_{\rm vir}$ \fi}
\def \rhocrit {\ifmmode \rho_{\rm crit} \else $\rho_{\rm crit}$ \fi}
\def \fbar {\ifmmode f_{\rm bar} \else $f_{\rm bar}$ \fi}

\def \rs {\ifmmode r_{\rm s} \else $r_{\rm s}$ \fi} 
\def \rh {\ifmmode r_{\rm h} \else $r_{\rm h}$ \fi} 
\def \Rvir {\ifmmode R_{\rm vir} \else $R_{\rm vir}$ \fi}
\def \Vvir {\ifmmode V_{\rm  vir} \else  $V_{\rm vir}$  \fi} 
\def \Vmax {\ifmmode V_{\rm  max} \else  $V_{\rm max}$  \fi} 
\def \Mvir {\ifmmode M_{\rm  vir} \else $M_{\rm  vir}$ \fi}  
\def \Nvir {\ifmmode N_{\rm  vir} \else $N_{\rm  vir}$ \fi}  
\def \Jvir {\ifmmode J_{\rm vir} \else $J_{\rm vir}$ \fi} 
\def \Evir {\ifmmode E_{\rm vir} \else $E_{\rm vir}$ \fi} 
\def \lam {\ifmmode \lambda  \else $\lambda$ \fi} 
\def \lamp {\ifmmode \lambda^{\prime} \else $\lambda^{\prime}$  \fi} 
\def \lampc {\ifmmode \lambda^{\prime}_{\rm c} \else
  $\lambda^{\prime}_{\rm c}$  \fi} 

\def \xoff {\ifmmode x_{\rm off} \else $x_{\rm off}$ \fi}
\def \rhorms {\ifmmode \rho_{\rm rms} \else $\rho_{\rm rms}$ \fi}
\def \qbar {\ifmmode \bar{q} \else $\bar{q}$ \fi}

\def \ri {\ifmmode r_{\rm i} \else $r_{\rm i}$ \fi} 
\def \rf {\ifmmode r_{\rm f} \else $r_{\rm f}$ \fi} 
\def \Mi {\ifmmode M_{\rm i} \else $M_{\rm i}$ \fi} 
\def \Mf {\ifmmode M_{\rm f} \else $M_{\rm f}$ \fi}

\def \Mb {\ifmmode M_{\rm b} \else $M_{\rm b}$ \fi} 
\def \Md {\ifmmode M_{\rm d} \else $M_{\rm d}$ \fi} 
\def \Mg {\ifmmode M_{\rm g} \else $M_{\rm g}$ \fi} 
\def \Rb {\ifmmode R_{\rm b} \else $R_{\rm b}$ \fi} 
\def \Rd {\ifmmode R_{\rm d} \else $R_{\rm d}$ \fi} 
\def \Rg {\ifmmode R_{\rm g} \else $R_{\rm g}$ \fi} 
\def \mgal {\ifmmode m_{\rm gal} \else $m_{\rm gal}$ \fi} 
\def \rj {\ifmmode {\cal R}_j \else ${\cal R}_j$ \fi} 
\def \lamgal {\ifmmode \lambda_{\rm gal} \else $\lambda_{\rm gal}$ \fi} 
\def \Vcirc {\ifmmode V_{\rm circ} \else $V_{\rm circ}$ \fi} 
\def \Vrot {\ifmmode V_{\rm rot} \else $V_{\rm rot}$ \fi} 
\def \Vopt {\ifmmode V_{\rm opt} \else $V_{\rm opt}$ \fi} 

\def \DeltaIMF {\ifmmode \Delta_{\rm IMF} \else $\Delta_{\rm IMF}$ \fi}

\def \VV {\ifmmode V_{\rm 2.2}/V_{200} \else $V_{2.2}/V_{200}$ \fi} 
\def \dvr {\ifmmode \partial_{\rm VR} \else $\partial_{\rm VR}$ \fi} 

\title[Dark halo response and halo quenching]{The response of dark matter haloes to elliptical galaxy formation: a new test for quenching scenarios.}

\author[Dutton et al.]  {Aaron  A.
  Dutton$^{1}$\thanks{dutton@mpia.de}, Andrea V. Macci\`o$^{1}$,
  Gregory S. Stinson$^1$, Thales A. Gutcke$^1$,  \newauthor{Camilla
    Penzo$^1$, Tobias Buck$^1$}\\ $^1$Max-Planck-Institut f\"ur
  Astronomie, K\"onigstuhl 17, 69117 Heidelberg, Germany}

\begin{document}
  
\date{accepted to MNRAS}
             
\pagerange{\pageref{firstpage}--\pageref{lastpage}}\pubyear{2015}

\maketitle           

\label{firstpage}
             

\begin{abstract}
  We use cosmological hydrodynamical zoom-in simulations with the SPH
  code {\sc gasoline} of four haloes of mass $M_{200}\sim
  10^{13}\Msun$ to study the response of the dark matter to elliptical
  galaxy formation. Our simulations include metallicity dependent gas
  cooling, star formation, and feedback from massive stars and
  supernovae, but not active galactic nuclei (AGN).  At $z=2$ the
  progenitor galaxies have stellar to halo mass ratios consistent with
  halo abundance matching, assuming a Salpeter initial mass function.
  However by $z=0$ the standard runs suffer from the well known
  overcooling problem, overpredicting the stellar masses by a factor
  of $\gta 4$.  To mimic a suppressive halo quenching scenario, in our
  forced quenching (FQ) simulations, cooling and star formation are
  switched off at $z=2$.  The resulting $z=0$ galaxies have stellar
  masses, sizes and circular velocities close to what is observed.
  Relative to the control simulations, the dark matter haloes in the
  FQ simulations have contracted, with central dark matter density
  slopes $d\log\rho/d\log r \sim -1.5$, showing that dry merging alone
  is unable to fully reverse the contraction that occurs at $z>2$.
  Simulations in the literature with AGN feedback however, have found
  expansion or no net change in the dark matter halo. Thus the
  response of the dark matter halo to galaxy formation may provide a
  new test to distinguish between ejective and suppressive quenching
  mechanisms.
\end{abstract}


\setcounter{footnote}{1}


\section{Introduction}
\label{sec:intro}

Dissipationless simulations run in the concordance \LCDM cosmology
make robust predictions for the structure of cold dark matter haloes
(e.g., Navarro \etal 1997, 2010; Bullock \etal 2001; Diemand \etal
2007; Macci\`o \etal 2007; Stadel \etal 2009; Zhao \etal 2009; Klypin
\etal 2011; Dutton \& Macci\`o 2014).  One of these predictions is
that CDM haloes should have ``cuspy'' central density profiles that
scale as $\rho(r) \propto r^{-\alpha}$ with $\alpha\approx
-1.2$. Observations have yet to unambiguously find these
cusps. Furthermore, a variety of studies based on gas dynamics,
stellar dynamics and gravitational lensing on scales from dwarf
galaxies to galaxy clusters often favor cored ($\alpha=0$) or shallow
cusps (i.e., $\alpha\lta 0.5$) instead (e.g., de Blok \etal 2001;
Swaters \etal 2003; Sand \etal 2004; Goerdt \etal 2006; Kuzio de Naray
\etal 2008; Walker \& Pe\~narrubia 2011; Oh \etal 2011; Newman \etal
2011; 2013).

This lack of observational evidence for cuspy dark matter density
profiles is often used as evidence against the CDM paradigm.  However,
a major complication in the comparison between observations of dark
matter density profiles and those predicted by cosmological N-body
simulations is the uncertain impact of baryons. Baryonic processes can
both increase and decrease the density profiles of dark matter haloes.
This currently prevents the observed structure of dark matter haloes
from being used as a robust test of the cold dark matter model.  A
fully predictive theory for the structure of CDM haloes must take into
account the effects of galaxy formation.

If the accretion of baryons onto the central galaxy is smooth and slow
then dark matter haloes should contract adiabatically (Blumenthal
\etal 1986). This process can increase the density of dark matter
haloes by an order of magnitude. Other processes can cause the dark
matter halo to expand, such as transfer of energy/angular momentum
from baryons to the dark matter via dynamical friction due to minor
mergers (El-Zant \etal 2001, 2004; Nipoti \etal 2004; Jardel \&
Sellwood 2009; Johansson \etal 2009; Lackner \& Ostriker 2010; Cole
\etal 2011;  Laporte \& White 2015), or galactic bars (Weinberg \&
Katz 2002; Sellwood (2008), but see McMillan \& Dehnen 2005), and
rapid mass loss/time variability of the potential due to supernovae
(SN) / stellar / active galactic nuclei (AGN) feedback (Navarro \etal
1996; Read \& Gilmore 2005; Mashchenko \etal 2006, 2008; Peirani \etal
2008; Governato \etal 2010; Pontzen \& Governato 2012; Macci\`o \etal
2012; Martizzi \etal 2012, 2013).  Each of these processes is likely
to occur during galaxy formation, but at present it is not clear which
process (if any) dominates, and on which galaxy mass scales.

Turning the problem around, if one can observationally measure the
halo response, then in the context of $\LCDM$, this gives clues to the
dominant galaxy formation mechanisms. In particular, as we discuss in
this paper, the halo response may provide a test for different galaxy
quenching models.  While it is known that most of the stars in massive
galaxies formed at high redshifts $z\gta 2$ (e.g., Thomas \etal 2005),
the mechanism(s) responsible for shutting down star formation
(quenching), and keeping it off (quiescence) are a subject of much
debate. A popular idea is that feedback of energy from the growth of
supermassive black holes (SMBH) ejects cold gas from galaxies and heats the
halo gas, preventing it from cooling onto the central galaxy. We refer
to this class of models as ``AGN quenching''. It is clear there is
enough energy released from the AGN to quench star formation, of issue
is whether the energy released by the AGN can couple efficiently to
the surrounding gas (e.g., Cielo \etal 2014).

An alternative idea is that cooling becomes very inefficient, and
effectively shuts down, once the halo mass reaches a critical mass of
$\sim 10^{12}\Msun$.  We refer to this class of models as ``halo
quenching''.  In practice the boundary between these two quenching
mechanisms is blurry because AGN can also act as a heat source for the
halo gas, although there are non-AGN heating sources such as young and
old stellar populations (Kannan \etal 2014, 2015; Conroy \etal
2015). The key difference in these mechanisms is that  AGN quenching
involves the episodic (and possibly violent) removal of gas from the
galaxy center, whereas halo quenching involves the suppression of gas
cooling from the halo.

Galaxy formation models implementing both mechanisms can broadly
reproduce the colors and mass functions of present day galaxies (e.g.,
Croton \etal 2006; Cattaneo \etal 2006; Schaye \etal 2015).  However,
there is at present no clear way to observationally distinguish
between these models. The correlation between SMBH mass and galaxy
properties is often cited as evidence for a physical link between
SMBHs and host galaxies, although (Jahnke \& Macci\`o 2011) showed
that such a correlation can naturally arise out of hierarchical
merging.

As discussed above, in terms of formation mechanisms, there is one
process that makes haloes contract: dissipative gas accretion
(Blumenthal \etal 1986), which must occur to form an elliptical
galaxy. Two processes have been investigated that make haloes expand:
mass outflows driven by AGN feedback (Martizzi \etal 2012); and
dynamical friction between baryons and dark matter during the galaxy
assembly process (El-Zant \etal 2001).  Dry mergers are likely to
be more important at late times, while AGN feedback is likely to be
more important at early times.

The goal of this study is to determine the impact of dissipationless
galaxy assembly on the structure of \LCDM haloes. In particular, we
wish to know whether dry merging can reverse the effects of halo
contraction that are expected to occur in the high redshift
progenitors.  To achieve this we use  fully cosmological
hydrodynamical simulations of the formation of $M_{200}\sim
10^{13}\Msun$ haloes using the smoothed particle hydrodynamics (SPH)
code {\sc gasoline}. We use the MaGICC (Stinson \etal 2013) star
formation and stellar feedback model which has been shown to produce
realistic disk galaxies (Brook \etal 2012) and successfully matches
the observed galaxy formation efficiencies across cosmic time in
haloes less massive than $\sim 10^{12}\Msun$ (Stinson \etal 2013; Wang
\etal 2015).  Our study improves on previous analytic and N-body
calculations (e.g., El-Zant \etal 2001; Lackner \& Ostriker 2010) in
that our mass assembly histories are fully cosmological, and compared
to previous cosmological hydro simulations (e.g., Johansson \etal
2009) our progenitor galaxies have realistic sizes and galaxy
formation efficiencies. 

This paper is organized as follows: the simulations including sample
selection, hydrodynamics, star formation and feedback models are
described in \S\ref{sec:simulations}. Results relating to global
parameters such as stellar masses, halo masses, galaxy sizes and
circular velocities are presented in \S\ref{sec:global}. The radial
mass profiles including the response of the dark matter to galaxy
formation are presented in \S\ref{sec:haloresponse}. We discuss
implications for our results in  \S\ref{sec:discussion} and give a
summary in \S\ref{sec:summary}.

\section{Simulations}
\label{sec:simulations}

The simulations presented here are fully cosmological ``zoom-in''
simulations of galaxy formation run in a flat \LCDM
cosmology. Cosmological parameters (see Table \ref{tab:cosmology}) are
based on the {\it Wilkinson Microwave Anisotropy Probe (WMAP)} 5th
year (Komatsu \etal 2009) and 7th year (Komatsu \etal 2011) results.
Haloes are selected from two parent dark matter only simulations run
with the {\sc{pkdgrav}} tree-code (Stadel 2001): a 90 Mpc box from
Macci\`o \etal (2008) using the WMAP5 cosmology, and a 80 Mpc$/h$ box
from Penzo \etal (2014) using the WMAP7 cosmology.

\begin{table}
 \centering
 \caption{Cosmological parameters. Column (1) cosmology ID. Column
   (2), $\Omega_{\rm m}$,  present day matter density. Column (3),
   $\Omega_{\Lambda}$, dark energy density. Column (4), $\Omega_{\rm
     b}$, baryon density. Column (5), $H_0$, Hubble parameter. Column
   (6), $\sigma_8$, power spectrum normalization. Column (7), $n$,
   power spectrum slope.}
  \begin{tabular}{ccccccccc}
\hline
\hline  
Cosmology & $\Omega_{\rm m}$ & $\Omega_{\Lambda}$ & $\Omega_{\rm b} $& $H_0$ & $\sigma_8$ & $n$\\
(1) & (2) & (3) & (4) & (5) & (6) & (7)\\
\hline
WMAP5 & 0.2580 & 0.7420 & 0.0438 & 0.720 & 0.796 & 0.963\\
WMAP7 & 0.2748 & 0.7252 & 0.0458 & 0.702 & 0.816 & 0.968\\
\hline
\hline
\label{tab:cosmology}
\end{tabular}
\end{table}

\begin{table*}
 \centering
 \caption{Simulation parameters. Column (1), name of initial
   conditions. Column (2), $M_{200}$, present day halo mass from
   adiabatic gas simulation. Column (3), $m_{\rm dark}$, dark matter
   particle mass. Column (4),  $m_{\rm gas}$, initial gas particle
   mass. Column (5), $\epsilon_{\rm dark}$, dark matter   particle
   force softening in comoving kpc. Column (6), $\epsilon_{\rm gas}$,
   gas (and star) particle force softening in comoving kpc. Column
   (7),  $N_{\rm dark}$, number of high-resolution dark matter
   particles. Column (8),  $N_{\rm gas}$, initial number of gas
   particles. Column (9), $n_{\rm th}$,  threshold for star
   formation. Column (10) name of cosmology -- see
   Table~\ref{tab:cosmology}. Column (11) symbol used in figures.}
  \begin{tabular}{lcccccrrccc}
\hline
\hline  
Name & $\log_{10}M_{200}$ & $\log_{10}m_{\rm dark}$ & $\log_{10}m_{\rm gas}$ & $\epsilon_{\rm dark}$ & $\epsilon_{\rm gas}$ & $N_{\rm dark}$ & $N_{\rm gas}$ & $n_{\rm th}$ & Cosmology & symbol\\ 
     & [M$_{\odot}$] & [M$_{\odot}$] & [M$_{\odot}$] & [ckpc] & [ckpc] & Million & Million & [cm$^{-3}$] \\
(1) & (2) & (3) & (4) & (5) & (6) & (7) & (8) & (9) & (10) & (11)\\
\hline
halo1   & 13.15 & 6.91 & 6.23 & 1.61 & 1.61$\phantom 0$ & 2.49 & 2.49 & 9.6$\phantom 0$& WMAP5 & triangle\\
halo2   & 13.19 & 6.91 & 6.23 & 1.61 & 1.61$\phantom 0$ & 2.49 & 2.49 & 9.6$\phantom 0$ & WMAP5 & square\\
halo3   & 13.20 & 6.91 & 6.23 & 1.61 & 1.61$\phantom 0$ & 2.49 & 2.49 & 9.6$\phantom 0$ & WMAP5 & pentagon\\
halo4.0 & 13.36 & 8.13 & 7.43 & 4.06 & 1.81$\phantom 0$ & 0.34 & 0.34 & 1.16 & WMAP7 & circle\\ 
halo4.1 & 13.35 & 7.23 & 6.53 & 2.04 & 0.910 & 2.68 & 2.68 & 1.16 & WMAP7 & circle\\
halo4.2 & 13.33 & 6.70 & 6.00 & 1.36 & 0.606 & 9.06 & 9.06 & 1.16 & WMAP7 & circle\\
halo4.3 & 13.42 & 6.32 & 5.63 & 1.02 & 0.455 &21.5$\phantom 0$ & 21.5$\phantom 0$ & 1.16 & WMAP7 & circle\\ 
\hline
\hline
\label{tab:setup}
\end{tabular}
\end{table*}

\subsection{Sample selection and initial conditions}
We select four haloes to re-simulate at higher resolutions.  The two
selection criteria are (1) a present day halo mass of $\sim
10^{13}\Msun$;  and (2) no large structures present within three
virial radii. Table ~\ref{tab:setup} lists the dark matter and baryon
particle masses and force softenings (in comoving kpc). To test for
numerical convergence, one of the haloes (halo4) has been run at four
different resolution levels, with particle masses varying by a factor
of 64, and force softenings by a factor of 4.  Note that for halo4 the
slightly higher halo mass in the highest resolution run (halo4.3) is
due to a merger event that occurs slightly earlier in this simulation
compared to the lower resolution runs.  

\begin{figure}
\centerline{
\psfig{figure=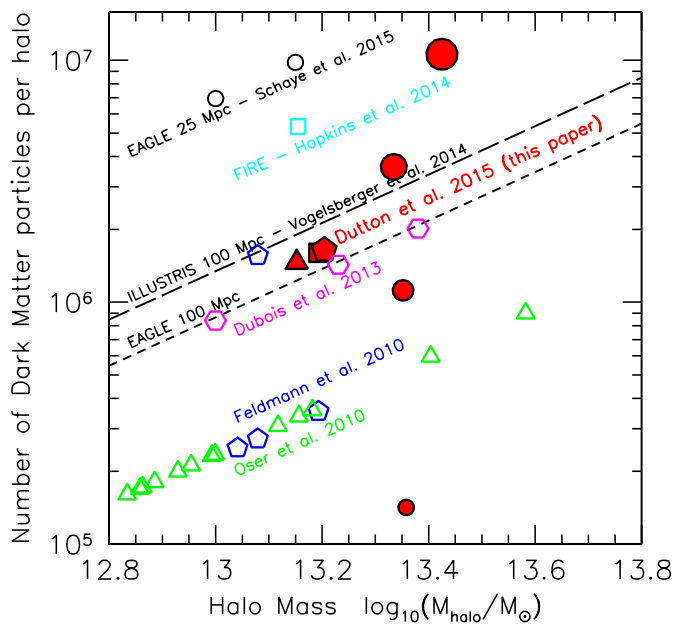,width=0.5\textwidth}
}
\caption{Dark matter mass resolution of our simulations (red filled
  symbols) compared to state-of-the-art simulations of elliptical
  galaxies in the literature (open symbols).}
\label{fig:massres}
\end{figure}

Fig.~\ref{fig:massres} shows the dark matter mass resolution of our
simulations (red filled symbols) compared to a number of
state-of-the-art simulations in the literature of comparable mass
haloes. We only consider simulations that have been run all the way to
redshift $z\simeq0$. We note that the simulations from Feldmann \&
Mayer (2015) are higher resolution, with a dark matter particle mass
of $7.9\times 10^5\Msun$, however, these simulations are only run to
$z=2$, where the halo mass is $\sim 3\times 10^{12}\Msun$ and thus
they do not appear in our comparison.

The highest resolution simulation of a $\sim 10^{13}\Msun$ halo in the
literature is from the 25 Mpc box from the EAGLES project (Schaye
\etal 2015, open circles). This simulation has a dark matter particle
mass of $m_{\rm DM}=1.21\times 10^{6}\Msun$ and thus $\sim 7$ million
particles for a $\sim 10^{13} \Msun$ halo. Comparable resolution can
be obtained with zoom-in simulations with a fraction of the
computational cost.  The FIRE project (Hopkins \etal 2014, cyan
square) includes a simulation with a dark matter particle mass of
$m_{\rm DM}=2.26\times10^6\Msun$, and roughly 5 million dark matter
particles within the virial radius.  For comparison our highest
resolution runs of halo4 have 3.6 and 10.6 million dark matter
particles inside $R_{200}$, and thus are among the highest resolution
elliptical galaxy simulations performed to date.

At the fiducial resolution our simulations have $\sim 1.5$ million
dark matter particles inside $R_{200}$. This is similar to that of the
high resolution halo in Feldmann \etal (2010, blue pentagon), the
haloes in Dubois \etal (2013, magenta hexagons),  together with the
large volume ($\sim 100$ Mpc) simulations by the ILLUSTRIS
(Vogelsberger \etal 2014) and EAGLE  collaborations (dashed lines).
Earlier ``zoom-in'' simulations from Oser \etal (2010) and Feldmann
\etal (2010) have particle masses of $\sim 3.6\times 10^{6}\Msun$, and
thus $\sim 300,000$ dark matter particles per halo.

\begin{figure*}
\centerline{
  \psfig{figure=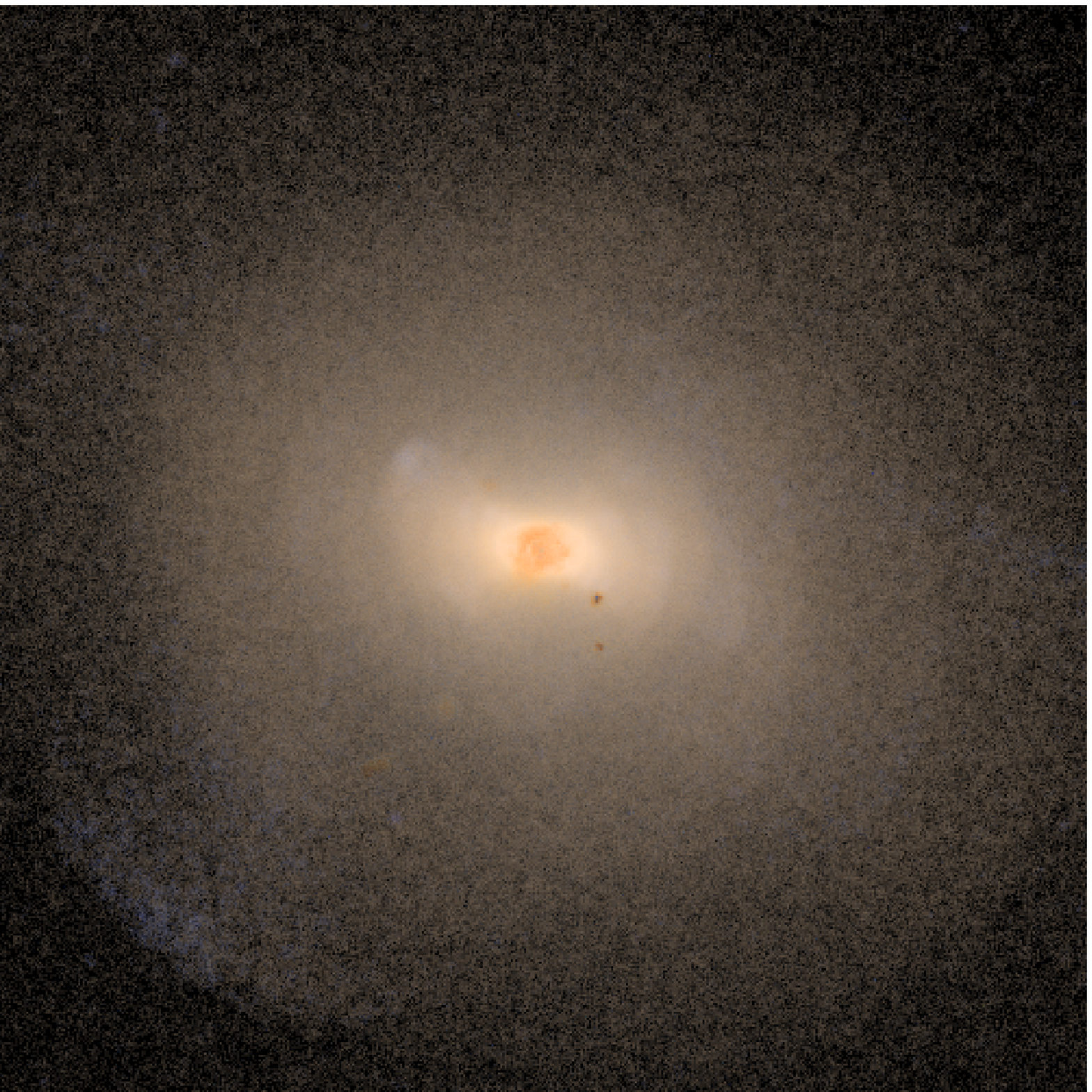,width=0.24\textwidth}
  \psfig{figure=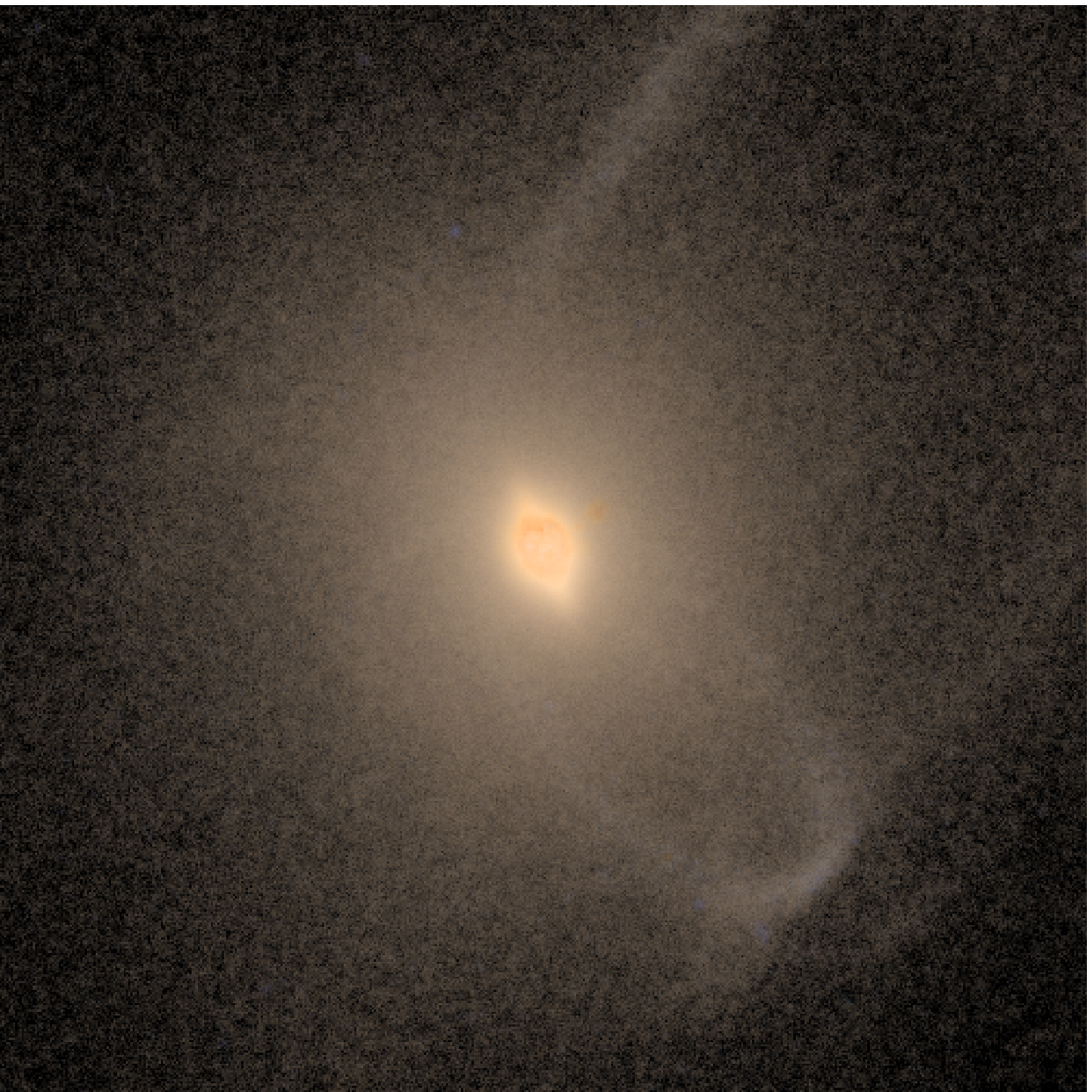,width=0.24\textwidth}
  \psfig{figure=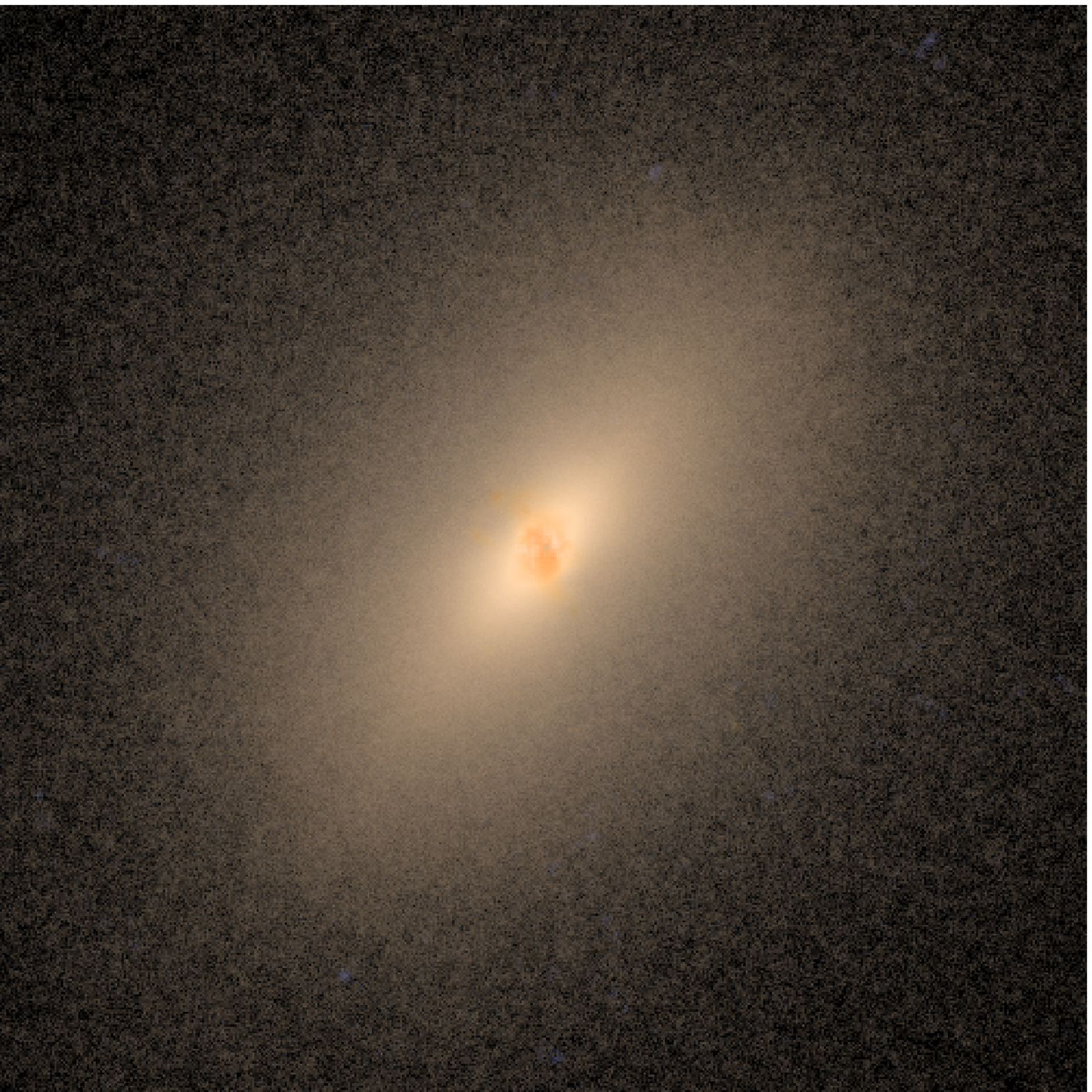,width=0.24\textwidth}
  \psfig{figure=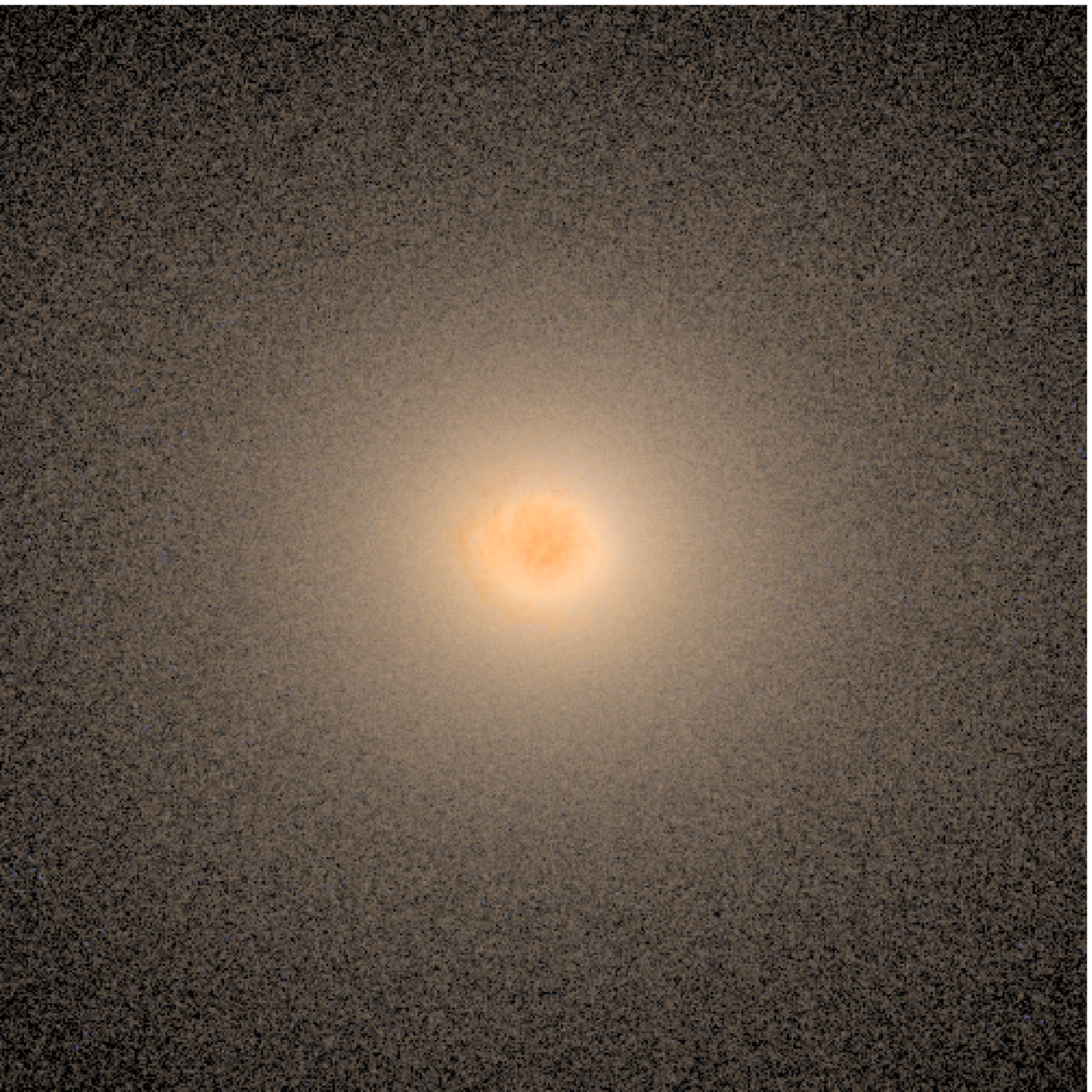,width=0.24\textwidth}
}
\centerline{
  \psfig{figure=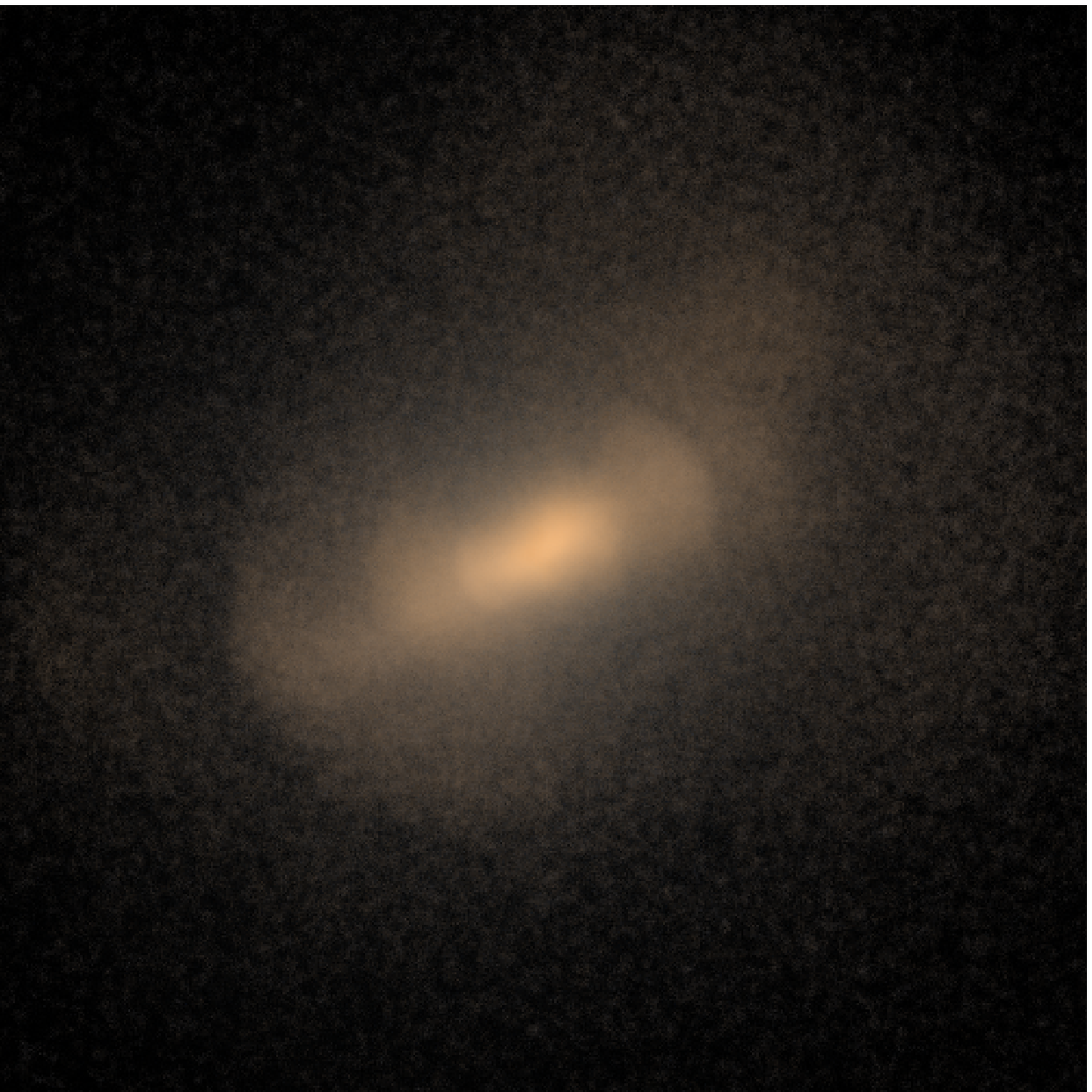,width=0.24\textwidth}
  \psfig{figure=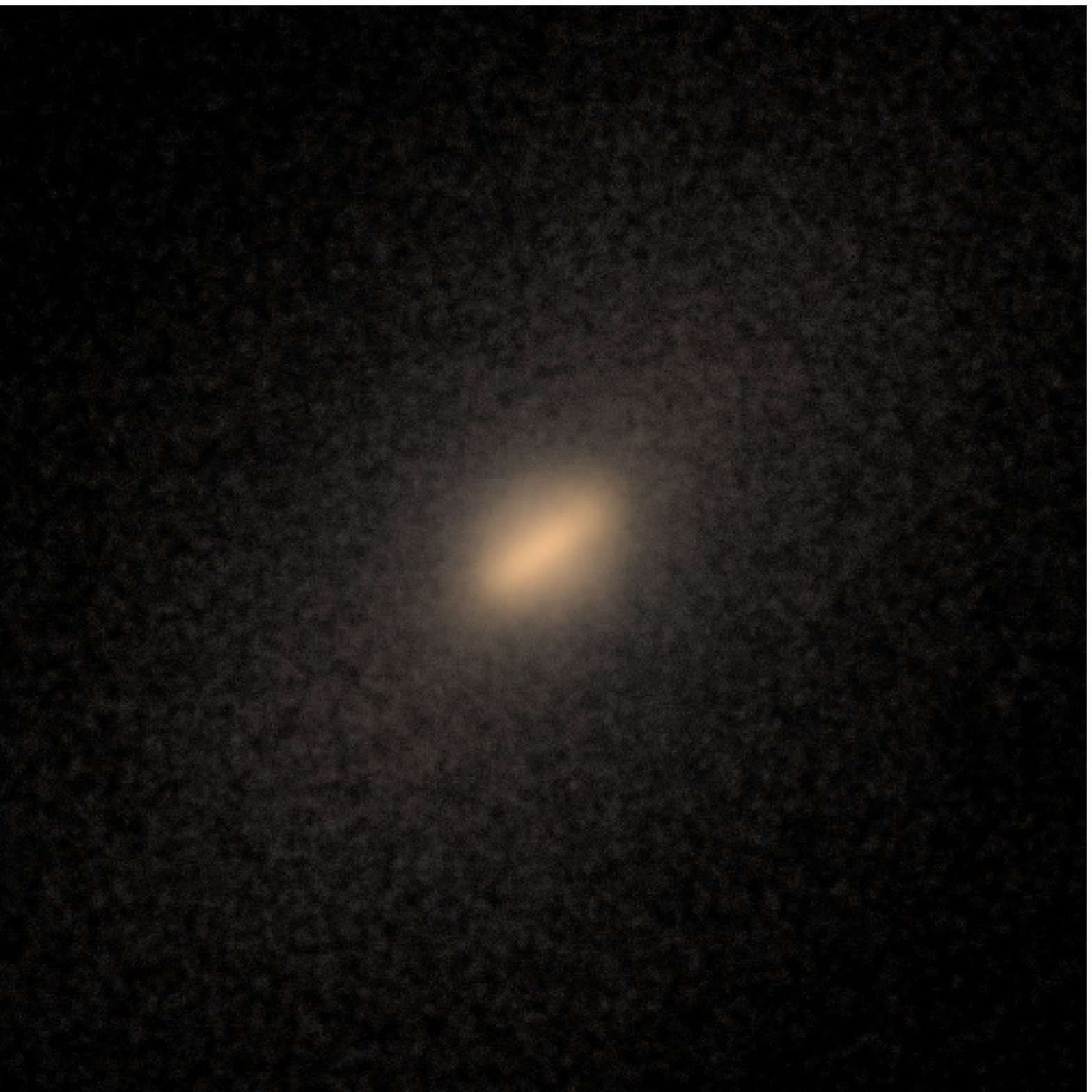,width=0.24\textwidth}
  \psfig{figure=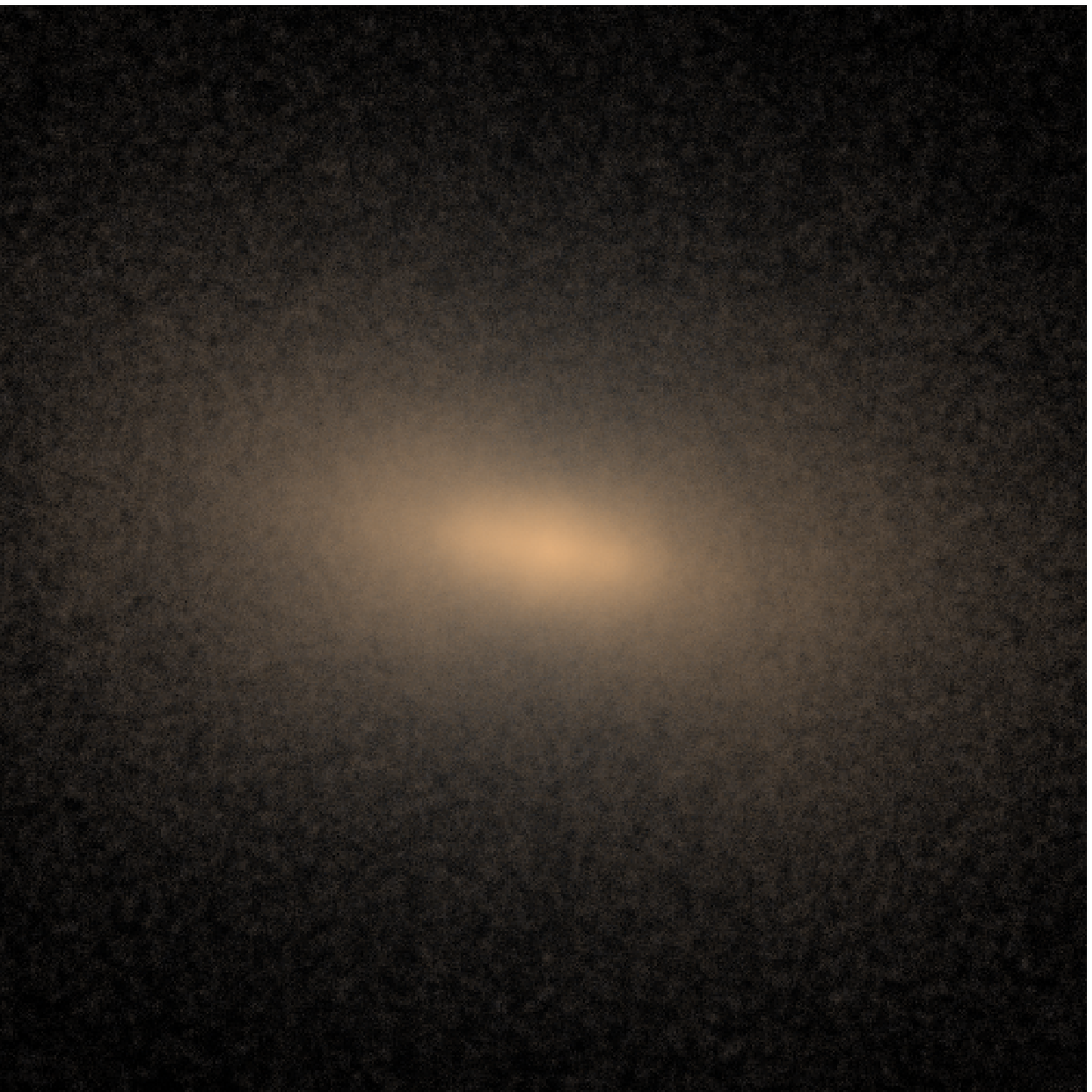,width=0.24\textwidth}
  \psfig{figure=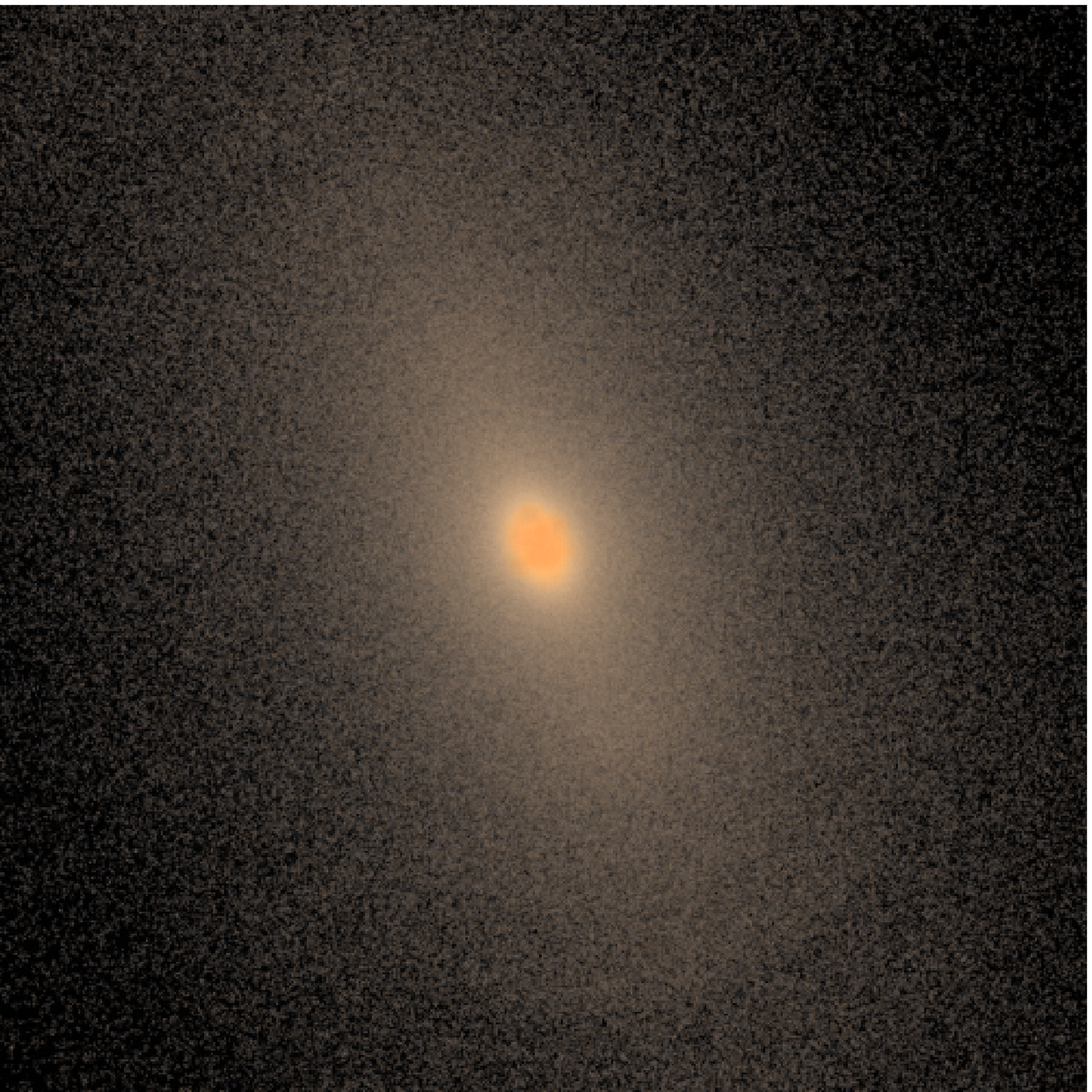,width=0.24\textwidth}
}
\caption{ {\sc{sunrise}} images of simulations at redshift $z=0$. Each
  box is 50 kpc on a side. Upper two panels show results for standard
  (overcooled) simulations, while lower two panels show forced
  quenching simulations. From left to right: halo1, halo2, halo3, and
  halo4.2.}
\label{fig:images}
\end{figure*}

\subsection{Hydrodynamics}
Our standard simulations use the same baryonic physics that was used
in the MaGICC project (see Stinson \etal 2013), based on the smoothed
particle hydrodynamics (SPH) code {\sc{gasoline}} (Wadsley \etal
2004).

Cooling via hydrogen, helium, and various metal-lines in a uniform
ultraviolet ionizing background is included as described in Shen \etal
(2010) and was calculated using cloudy (version 07.02; Ferland \etal
1998). These calculations include photo ionization and heating from
the Haardt \& Madau (2005) UV background and Compton cooling in the
temperature range 10 to $10^9$ K. In the dense, interstellar medium
gas, we do not impose any shielding from the extragalactic UV field as
the extragalactic field is a reasonable approximation in the
interstellar medium. Diffusion of metals and thermal energy between
particles has been implemented as described in Wadsley \etal (2008).

Stars form from cool dense gas that has reached a temperature of
$T<1.5 \times 10^4$K and a density of  $n> n_{\rm th}$. For halo1-3 we
adopt $n_{\rm th}=9.6$ cm$^{-1}$ following MaGICC (Stinson \etal
2013). For halo4 we adopt $n_{\rm th}=1.16$ cm$^{-1}$, which is a
conservative estimate of the maximum gas density that can be resolved
and is calculated using $n_{\rm th}=32 (m_{\rm gas}/5)/\epsilon_{\rm
  gas}^3$, where $m_{\rm gas}$ is the initial gas particle mass,
$m_{\rm gas}/5$ is the minimum gas particle mass, and $\epsilon_{\rm
  gas}$ is the gas force softening. The factor 32 is the number of SPH
smoothing elements.  This difference in star formation threshold has
no noticeable impact on the structure of the galaxies or dark matter
halo.
Star formation follows the Kennicutt-Schmidt law with
10 percent efficiency of turning gas into stars during one dynamical
time (Stinson \etal 2006). The stellar mass distribution in each star
particle follows the Chabrier (2003) stellar initial mass function
(IMF).

Massive stars explode as Type II SN and deposit an energy of $E_{\rm
  SN} = 10^{51}$ erg into the surrounding gas. Cooling for gas
particles subject to SN feedback is delayed based on the subgrid
approximation of a blast wave as described in Stinson \etal (2006).
Furthermore, radiation energy from massive stars is considered since
molecular clouds are disrupted before the first SN explosion (which
happens after 4 Myr from the formation of the stellar population). We
assume that 10 percent of the total radiation energy is coupled with
the surrounding gas. The inclusion of this early stellar feedback
(ESF) reduces star formation before SN start exploding. Thus, after
the ESF heats the gas to $T > 10^6$ K, the gas rapidly cools to $10^4$
K, which creates a lower density medium than if the gas were allowed
to continue cooling until SN exploded.  Stinson \etal (2013) and
Wang \etal (2015) show how this feedback mechanism limits star
formation to the amount prescribed by halo abundance matching (e.g.,
Behroozi \etal 2013) at all redshifts for  haloes of present day
virial mass $\sim 10^{10} - 10^{12}\Msun$.

For each initial condition we run three simulations that differ only in their
treatment of cooling and star formation.
\begin{itemize}
\item {\bf Standard:} cooling, star formation, and stellar feedback at
  all times following MaGICC (Stinson \etal 2013). Feedback from
    AGN is not included.
\item {\bf Forced Quenching (FQ):} same as standard down to $z\simeq2.1$,
  after which  the simulation is evolved with adiabatic gas until
  $z=0$ -- i.e., there is no further star formation.
\item {\bf No Cooling:} Gas is adiabatic at all times, and thus does
  not become cool and dense enough to form stars. These simulations
  act as a control for the effects of galaxy formation (cooling, star
  formation, and feedback) on the dark halo structure. 
\end{itemize}

The forced quenching simulations act as a limiting case where cooling
is shut down and star formation stops rapidly. It allows us to study
the effects of dry merging on the structure of the galaxy and dark
matter halo with cosmologically consistent initial conditions and
merger histories. As we show below this simple prescription allows one
to reproduce the stellar mass vs halo mass relation from abundance
matching, which is not possible when adopting standard metal line
cooling, and no additional heating sources (such as AGN feedback). We
note that previous simulations that come close to matching the stellar
mass vs halo mass relation without AGN feedback (Oser \etal 2010;
Feldmann \etal 2010; Feldmann \& Mayer 2015) do so because they adopt
primordial metallicity gas in the calculation of the cooling
rate. Including realistic halo gas metallicities increases the cooling
rate by a factor of $\sim 2-3$ (e.g., Dutton \& van den Bosch 2012),
and thus would result in overcooled galaxies.  It has been shown that
ionizing radiation from young and old stellar populations can
significantly reduce cooling rates in $10^{12}\Msun$ haloes (Kannan
\etal 2014).  Additionally Conroy \etal (2015) argue that heating
provided by the winds of dying low-mass stars is capable of
suppressing cooling of hot gas for a Hubble time in  halo masses above
$\sim 10^{12.5}\Msun$ at redshifts below $z\sim 2$.  Thus, there is
motivation for studying the impact of inefficient gas cooling
on elliptical galaxy formation.

\subsection{Derived galaxy and halo parameters}
Haloes in our zoom-in simulations were identified using the MPI+OpenMP
hybrid halo finder \texttt{AHF}\footnote{http://popia.ft.uam.es/AMIGA}
(Knollmann \& Knebe 2009; Gill \etal 2004). \texttt{AHF} locates local
over-densities in an adaptively smoothed density field as prospective
halo centers. The virial masses of the haloes are defined as the
masses within a sphere containing $\Delta=200$ times the cosmic
critical matter density.  The virial mass, radius, and circular
velocity are denoted $M_{200}$, $R_{200}$, and $V_{200}$. At $z=0$
there is, by construction, one central halo in the zoom-in
region. However, at $z\sim 2$ there are several well resolved
progenitor galaxies which we also consider in the global parameter
evolution plots.

The mass in stars, $M_{\rm star}$, is measured within a sphere of
radius, $r_{\rm gal}\equiv0.2R_{200}$.  The stellar half-mass radius,
$r_{1/2}$, encloses within a sphere half of the stellar mass within
$r_{\rm gal}$. The half-mass circular velocity is defined as
$V_{1/2}\equiv \sqrt{G M(r_{1/2})/r_{1/2}}$, where $M(r_{1/2})$ is
the total mass within a sphere of radius $r_{1/2}$.  The (mass
weighted) slope of the total mass density profile, $\gamma'$, and the
``inner'' slope of the dark matter density profile, $\alpha$, are
measured between 0.01 and 0.02 $R_{200}$. The choice of this scale is
motivated by the following: it is used by previous studies (e.g., Di
Cintio \etal 2014); it is resolved in our simulations; and it
corresponds to the average half-light radii of galaxies (Kravtsov
2013) and is thus observationally accessible.

Images of our simulated galaxies at $z=0$ are shown in
Figs.~\ref{fig:images}.  Each image is 50 kpc on a side and was
created using the Monte Carlo radiative transfer code {\sc{sunrise}}
(Jonsson 2006).  The image brightness and contrast are scaled using
arcsinh as described in Lupton \etal (2004).  All simulated galaxies
have red colors and elliptical like morphology. Some show signs of
recent mergers in the form of shells and streams. The main apparent
difference between the standard (upper panels) and forced quenching
(lower panels) simulations is a reduction in surface brightness (the
same scale is used in all images). As is visible, and we show
quantitatively below, the half-light sizes are roughly the same in
each type of simulation (standard vs FQ).

\begin{figure}
\centerline{
\psfig{figure=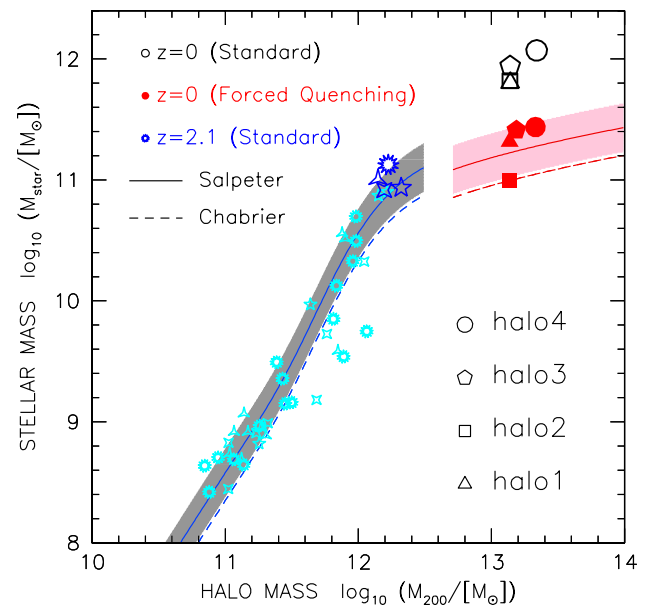,width=0.5\textwidth}  
}
\caption{Stellar mass vs halo mass relation for simulations at
  redshift $z=0$ and the progenitor galaxies at $z=2.1$. Open symbols
  show standard simulations (with cooling and star formation at all
  redshifts) at $z=0$ (black) and $z=2.1$ (blue for most massive
  progenitors, cyan for other progenitors). Solid red symbols show
  forced quenching simulations (which have no cooling or star
  formation since $z=2.1$). The dashed lines show the abundance matching
  relations from Behroozi \etal (2013) at redshifts $z=0$ (red) and
  $z=2.1$ (blue), the solid lines show the corresponding relations for
  stellar masses higher by 0.23 dex, corresponding to a Salpeter
  IMF. The shaded ares shows the $1\sigma$ scatter in stellar mass at
  fixed halo mass.}
\label{fig:mm}
\end{figure}

\section{Global Properties}
\label{sec:global}

Before we discuss the response of dark matter haloes to galaxy
formation in \S~\ref{sec:haloresponse}, we present several global
properties of our simulated galaxies and compare them to observations.
At $z=2.1$, the standard simulations compare well with observed
galaxies.  However, once they are evolved to $z=0$, the standard
simulations have too many stars, are too dense, and have mass profiles
that are too centrally concentrated.  Thus, they are poor candidates
for studying halo contraction.  In contrast, we show that the forced
quenching simulations share many properties with observed galaxies at
$z=0$, so it is more interesting to study them as templates for
elliptical galaxy formation. 

\begin{figure*}
\centerline{
 \psfig{figure=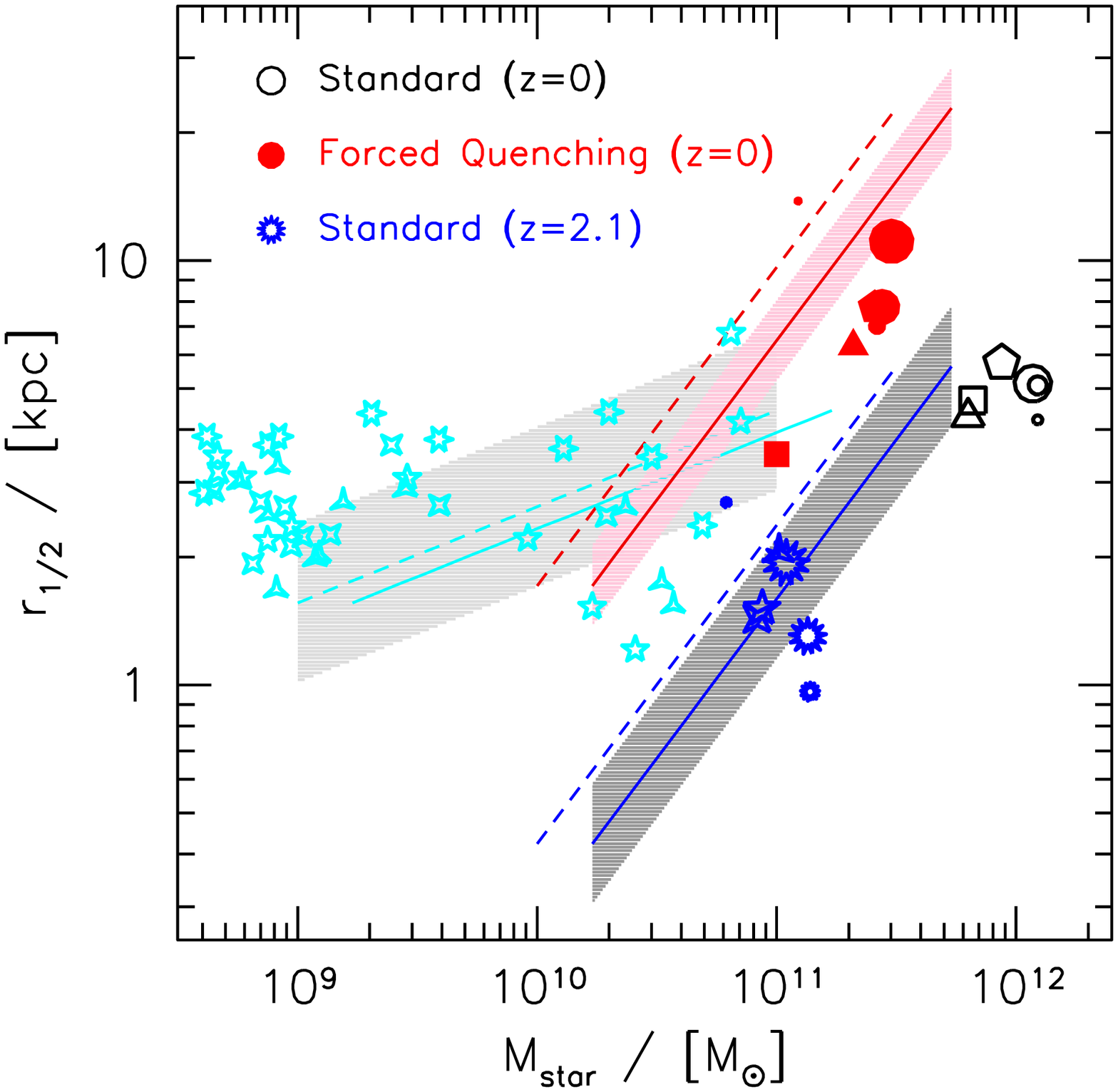,width=0.5\textwidth} 
\psfig{figure=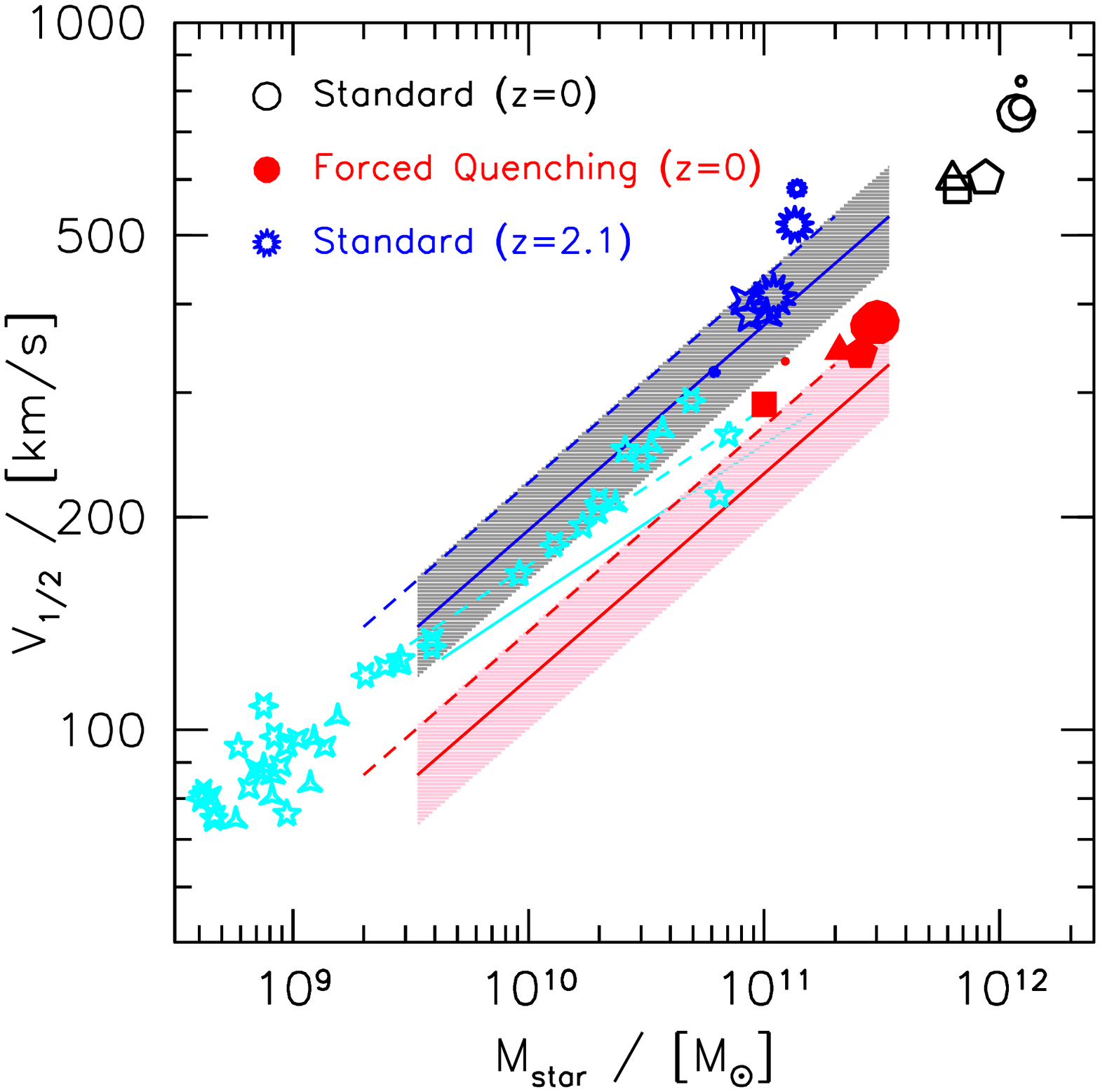,width=0.5\textwidth} 
}
\caption{Scaling relations between galaxy size, $r_{1/2}$, circular
  velocity at the half-mass size $V_{1/2}$, and galaxy stellar masses,
  $\Mstar$, at $z=2.1$ and $z=0$.  The observed relations for quiescent
  galaxies are shown at $z=0$ (red/pink) and $z=2.1$ (blue/grey) for
  both Chabrier (dashed) and Salpeter (solid) IMFs. The shaded regions
  show the $1\sigma$ scatter.  At $z=0$ the standard simulations are
  shown with black open symbols, and the forced quenching simulations
  are shown with red filled symbols.  At $z=2.1$ the most massive
  progenitors are shown with blue symbols, and the lower mass
  progenitors with cyan. The cyan lines show the observed size-mass
  and velocity-mass relations for star forming galaxies at $z=2.1$.}
  \label{fig:rmvm}
\end{figure*}

\subsection{Stellar mass vs halo mass}
One of the most fundamental properties of a galaxy or dark matter halo
is its mass. The relation between the mass in stars, $\Mstar$, and the
virial mass, $M_{200}$, of our simulated galaxies is shown in
Fig.~\ref{fig:mm}. Results are shown at redshift $z=0$ for standard
(black open symbols) and forced quenching (red filled symbols) and at
$z=2.1$ (blue open symbols show the most massive progenitor, while
cyan open symbols show other progenitors containing more than 20,000
total particles).  The lines show the relations from halo abundance
matching from Behroozi \etal (2013), with the shaded region showing
the $1\sigma$ intrinsic scatter. Two assumptions for the observed
stellar masses are shown: A Chabrier (2003) initial mass function
(IMF, dashed) as found in the Milky Way; and Chabrier+0.23 dex
(solid), which corresponds to  a Salpeter (1955) IMF and is likely a
better approximation for the centers of massive elliptical galaxies
(e.g., Conroy \& van Dokkum 2012; Dutton \etal 2013b).

At redshift $z=0$ the standard simulations over predict the Chabrier
IMF stellar masses  by an order of magnitude --- this is the classic
overcooling problem. Even if one adopts a Salpeter IMF, the stellar
masses are still over predicted by a factor of $\sim 4$.  Given the
$1\sigma$ intrinsic scatter in the stellar masses at fixed halo masses
is $\sim 0.2$ dex (More \etal 2011; Reddick \etal 2013), real galaxies
offset from the median relation by a factor of $\sim 4$ will be
extremely rare, if they even exist. Thus there is no escape from the
conclusion that these simulations have formed too many stars. At
$z=2.1$, however, the standard simulations have only a factor of $\sim
2$ too many stars compared to a Chabrier IMF.  Thus when adopting an
observed Salpeter IMF the simulations provide a good match to the halo
abundance matching results. Note that while there is a formal
inconsistency here between the IMFs used in our simulations (Chabrier)
and observation (Salpeter), in practice the feedback efficiency
parameters in the simulations can be simply rescaled.

When we disable cooling and force quenching at redshift $z=2.1$  the
resulting stellar masses at $z=0$ are consistent with halo abundance
matching constraints, provided we also adopt a Salpeter-like IMF at
$z=0$.  Thus, our simulations show that no new stars are needed to be
formed since redshift $z\sim 2$ in haloes of present day mass of
$M_{200} \sim 10^{13} \Msun$.  This result supports a two phase galaxy
formation picture (e.g., Oser \etal 2010). At halo masses below
$M_{200} \sim 10^{12}\Msun$ central galaxies form the majority of
stars {\it in situ}, with the efficiency being regulated by stellar
feedback. Above $M_{200}\sim 10^{12}\Msun$ in situ star formation is
suppressed, and the growth in stellar mass is dominated by
dissipationless mergers.

A consequence of the dissipationless assembly is that whatever IMF
characterizes the $z=0$ galaxies must also describe their most massive
progenitors. In our simulations progenitor galaxies with $\Mstar \gta
10^{10}\Msun$ at $z=2.1$ contribute 90\% of the present day mass.  Since
the lower mass progenitors ($\Mstar \lta 10^{10}\Msun$) don't
contribute significantly to the $z=0$ central galaxy mass, this
self-consistency argument does not constrain their IMF.  The stars
from the lower mass progenitors preferentially end up at large radii,
which would result in a radial IMF gradient if the low mass galaxies
have normal IMFs.

\begin{figure*}
\centerline{
  \psfig{figure=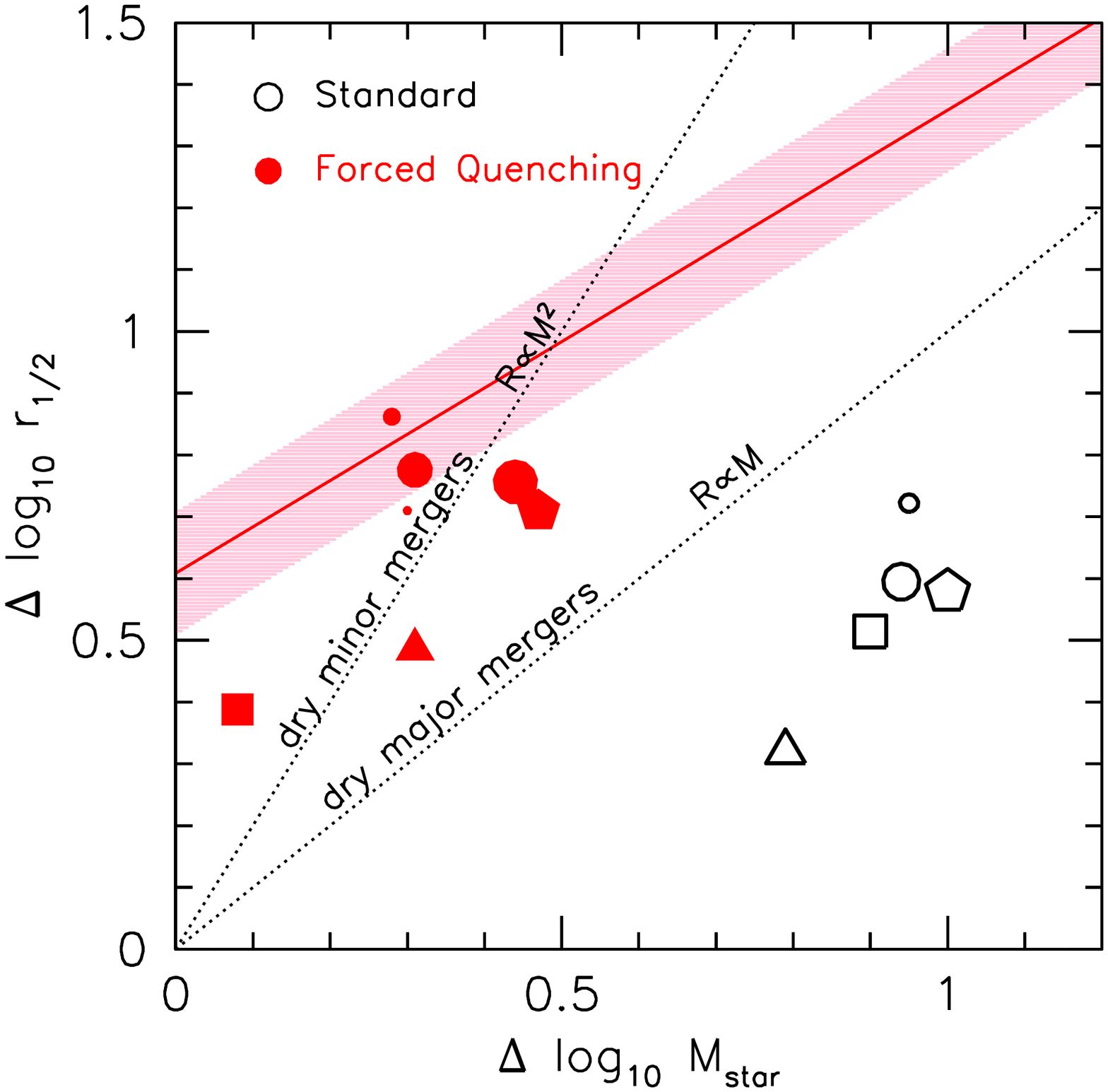,width=0.5\textwidth}
 \psfig{figure=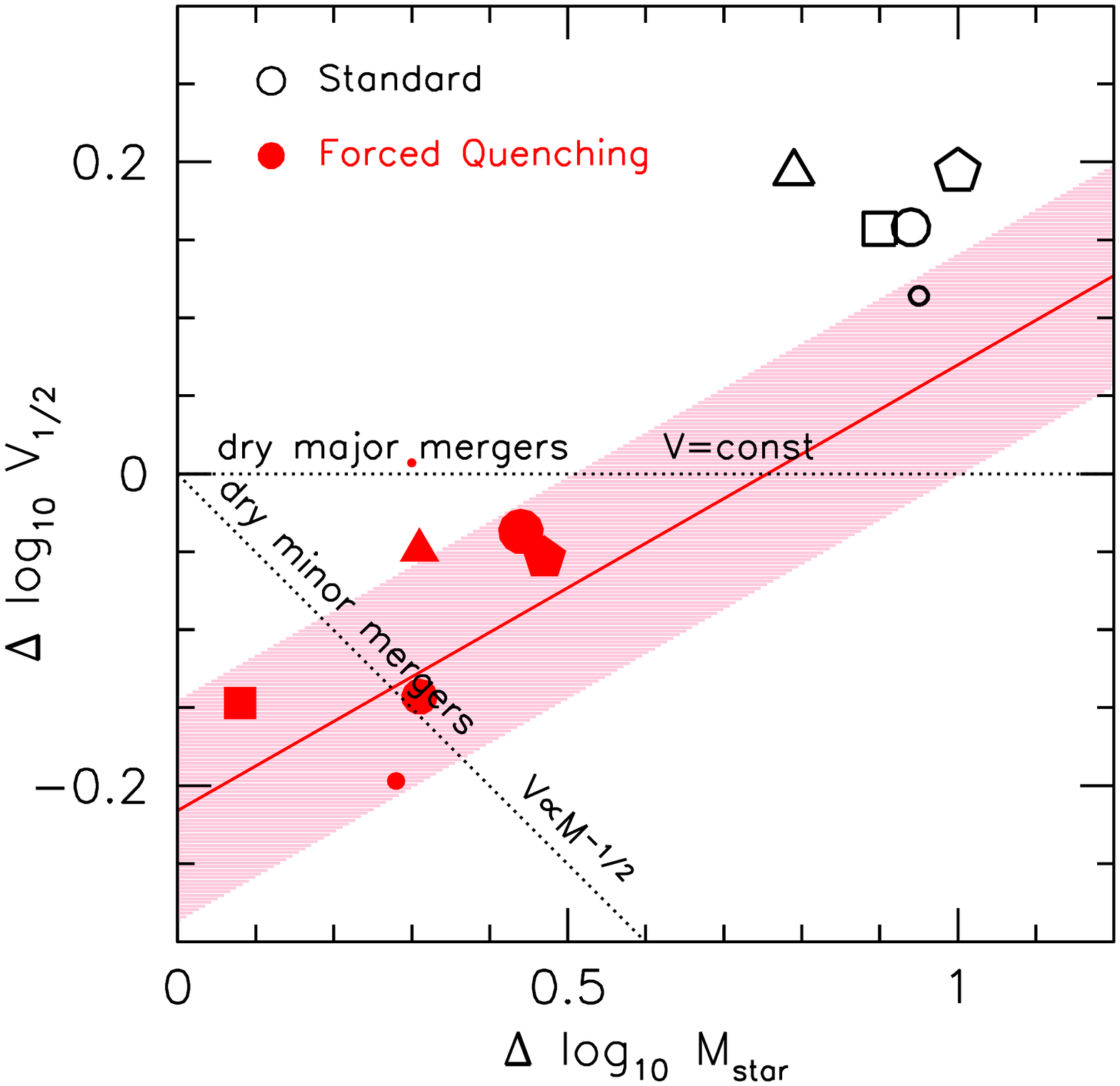,width=0.5\textwidth} 
}
\caption{Change in galaxy sizes, circular velocities, and stellar
  masses, between $z=2.1$ and $z=0$, of the most massive progenitor
  for the forced quenching (red, filled symbols) and standard (black,
  open symbols) simulations. The left panel shows the change in size
  vs change in mass, while the right panel shows the change in
  velocity vs change in mass. The red lines show the evolution
  required to stay on the size-mass and velocity-mass relations, with
  the shaded region showing the scatter in the $z=0$ relation. For
  reference, the dotted lines show the approximate evolutionary paths
  for dry major and minor mergers.}
\label{fig:drv}
\end{figure*}

\subsection{Galaxy sizes and circular velocities}
The next global properties we consider are the half-mass size of the
stellar mass distribution, $r_{1/2}$, and the circular velocity at
$r_{1/2}$. Fig.~\ref{fig:rmvm} shows the size-mass (left) and
velocity-mass (right) relations at $z\simeq 0$ and $z\simeq 2$.

For the size-mass relation we compare to observations from van der Wel
\etal (2014) --- who measure 2D major axis half-light radii in rest
frame V-band light for both quiescent and star forming galaxies.  The
observed relations for non-star-forming galaxies at $z=2.1$ and $z=0$
are shown with blue and red lines, respectively, with the shaded
region showing the $1\sigma$ intrinsic scatter. As in
Fig.~\ref{fig:mm} the solid lines assume a Salpeter IMF, while the
dashed lines assume a Chabrier IMF. At $z=2.1$ we also show the
size-mass relation for star forming galaxies with cyan lines.

In our simulations we measure spherical 3D half-stellar mass sizes. To
correct for projection effects we multiply the observed sizes of ETGs
by 4/3. No correction is applied to LTGs as the high disk fractions
suggest a minimal difference between projected and spherical aperture
sizes.  We leave an investigation of projection effects, mass-to-light
ratio variations and axis ratios to a future study. The purpose of our
current comparison is to determine if the simulated galaxies have
reasonable structural properties.  The observed relation between 3D
half-light size, stellar mass (assuming a Chabrier IMF) and redshift
is thus
\begin{equation}
  \frac{R_{1/2}}{[\kpc]}
  =  5.73\left(\frac{\Mstar}{[5\times10^{10}\Msun]}\right)^{0.75}  H(z)^{-1.29}.
\end{equation}

For the velocity-mass relation we use the $z=0$  Faber \& Jackson
(1976) relation for quiescent galaxies from Gallazzi \etal (2006). We
convert velocity dispersions into circular velocity using $V_{1/2}=1.5
\sigma_{\rm e}$ (Dutton \etal 2011b; Cappellari \etal 2013).  For the
evolution we adopt the scaling of $\sigma(\Mstar) \propto
(1+z)^{0.44}$ from the observed compilation of Oser \etal (2012). The
observed relation between circular velocity, stellar mass (assuming a
Chabrier IMF) and redshift is thus
\begin{equation}
\frac{V_{1/2}}{[\kms]} =  220\left(\frac{\Mstar}{[5\times10^{10}\Msun]}\right)^{0.29}  (1+z)^{0.44}.
\end{equation}
For star forming galaxies galaxies at $z\sim 2$ we use the
velocity-mass relation   from Dutton \etal (2011a) using data from
Cresci \etal (2009).

The most massive galaxies at $z=2.1$ in the standard simulations (blue
symbols) have half-mass sizes of $\sim 1-2$ kpc and circular
velocities of $\sim 400-500 \kms$, consistent with the size-mass and
velocity-mass relations of quiescent galaxies (dark grey shaded
regions).  The lower mass progenitors (cyan symbols) lie close to the
corresponding relations for star forming galaxies (shown with cyan
lines).  Thus the standard simulations form a realistic population of
progenitor galaxies, adding to the  successes of MaGICC feedback model
(Stinson \etal 2013) in reproducing global scaling relations of spiral
galaxies (Brook \etal 2012; Wang \etal 2015).

By $z=0$ in both the standard and FQ simulations the most massive
galaxies have grown substantially in size and mass. The standard
simulations (black open symbols) have grown roughly along the $z=2.1$
size-mass relation and are thus too dense compared to $z=0$
observations. The FQ simulations fall parallel to the observed
size-mass relation, being offset on average by factor of $\sim 1.5$.
Observational effects that may contribute to this  discrepancy are
discussed below. Here we note that numerical resolution may play a
role, as the sizes of halo4 have not shown convergence when we
increase the mass resolution by a factor of 8 (from level1 to
level3). By contrast, stellar masses are stable for simulations with
more than a million particles.

The changes since $z=2.1$ in the sizes, circular velocities, and
stellar masses are more clearly shown in  Fig.~\ref{fig:drv}.  For
both standard and FQ simulations the sizes increase on average by
$\sim 0.6$ dex, albeit with more variation in the FQ simulations. This
is interesting given that the mass evolution differs by a factor of
$\sim 4$ between standard and FQ simulations. However, the velocity
evolution is different: a decrease of $\sim -0.1$ dex for the FQ
simulations, and an increase of $\sim 0.15$ dex for the standard
simulations.

The dotted lines in Fig.~\ref{fig:drv} show the changes expected for
dry major ($R\propto M, V={\rm const}$) and minor mergers ($R\propto
M^2, V\propto M^{-1/2}$). (e.g., Bezanson \etal 2009; Hopkins \etal
2010). All but one simulation (halo2, squares) falls within these
expected scalings. In halo2 the size increases by a factor of 2.5 with
very little increase in stellar mass which points towards an adiabatic
expansion scenario. 

\begin{figure}
\centerline{
\psfig{figure=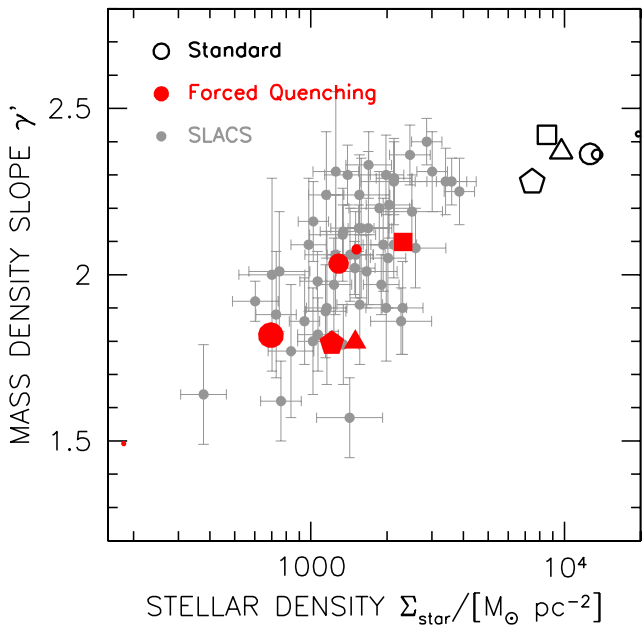,width=0.5\textwidth}
}
\caption{Logarithmic slope of the total mass profile between 1 and 2\%
  of the virial radius, $\gamma'$ vs the stellar surface density
  within the effective radius, $\Sigma_{\rm star}$. Points with error
  bars show observations from the SLACS survey (Auger \etal
  2010b). Standard simulations are shown with black open symbols,
  while forced quenching simulations are shown with red solid
  symbols.}
\label{fig:gammap}
\end{figure}

The solid lines in Fig.~\ref{fig:drv}  show the change in size and
velocity required to remain on the size-mass and velocity-mass scaling
relations between $z=2.1$ and $z=0$.  For no mass change, sizes need to
increase by 0.61 dex and velocities need to decrease by $-0.22$ dex. If
the stellar masses increase, even stronger evolution is required to
remain on the size-mass relation.  Our FQ simulations have the
observed amount of evolution in velocities, but don't have as much
evolution in sizes as observed, they are still too small by a factor
of $\sim 1.5$. There are several considerations to keep in mind when
interpreting the small size:

{\it 1) Population growth.} If there is substantial growth in the
number of quenched galaxies  at a given mass, then the scaling
relation will be dominated by the new arrivals. Thus earlier quenched
galaxies could be significantly offset from the relation without
violating the scatter.

{\it 2) Progenitor size bias.}  Star forming galaxies are observed to
be larger than quenched galaxies. If the quenching process does not
change the galaxy size, then  galaxies that are added to the quenched
population will have, on average, larger sizes than existing quenched
galaxies of the same mass. Thus the earlier quenched galaxies will not
need to grow as much in size as the population appears to.

{\it 3) M/L variations.} In the simulations we are measuring sizes in
stellar mass, while observations are for optical stellar
light. Galaxies are known to have smaller sizes in longer wavelengths
(e.g., van der Wel \etal 2014). For quiescent galaxies the implied
conversion from rest frame V-band to K-band (which presumably traces
stellar mass) is on average 0.15 dex.  Star forming galaxies have
stronger gradients at later times, which results in stronger evolution
in half-light than half-mass sizes by 0.14 dex. A similar effect for
quiescent galaxies would thus explain the entire discrepancy between
our simulations and observations.   In addition to color gradients,
IMF gradients could plausibly make sizes appear larger in light than
they are in mass. For example, if the most massive $z=2.1$ progenitor
formed with a uniformly heavy IMF, and then grew by adding galaxies
with lighter IMF to the outer parts. We note there are recent
observational hints of such radial IMF variations in massive quiescent
galaxies (Mart{\'{\i}}n-Navarro \etal 2015).

\begin{figure}
\centerline{
\psfig{figure=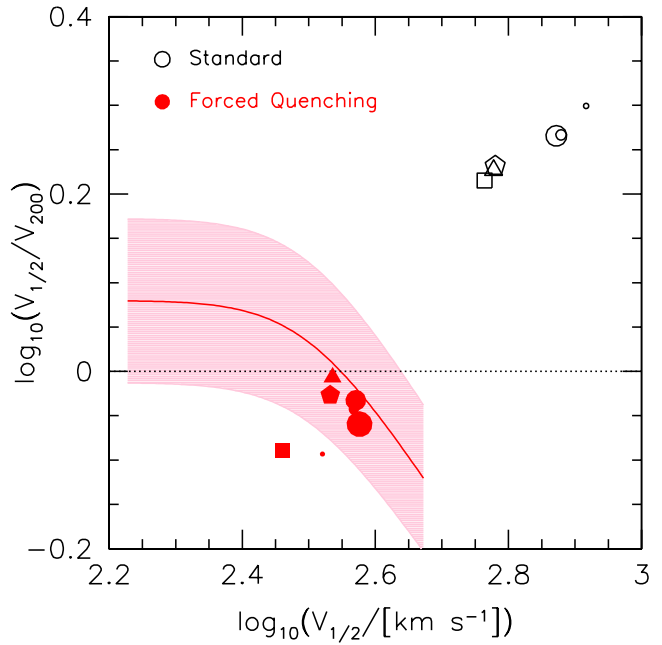,width=0.5\textwidth}
}
\caption{Relation between circular velocity at the half-mass radius,
  $V_{1/2}$, and virial radius, $V_{200}$ at $z=0$. Forced quenching
  simulations are shown with red solid points, and standard
  simulations with black open points. The shaded region shows
  observations from Dutton \etal (2010).}
\label{fig:vv}
\end{figure}

\begin{figure*}
\centerline{
  \psfig{figure=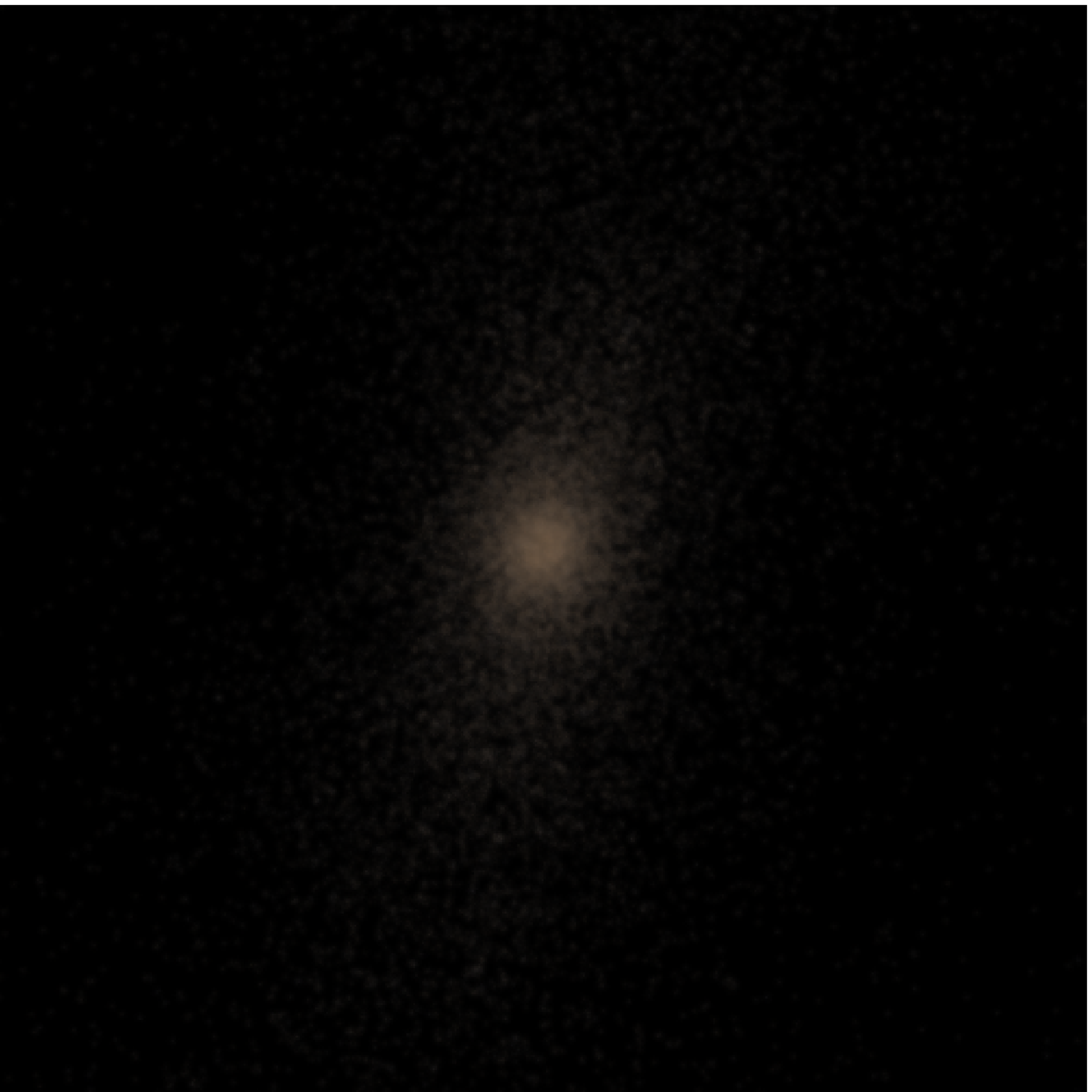,width=0.24\textwidth}
  \psfig{figure=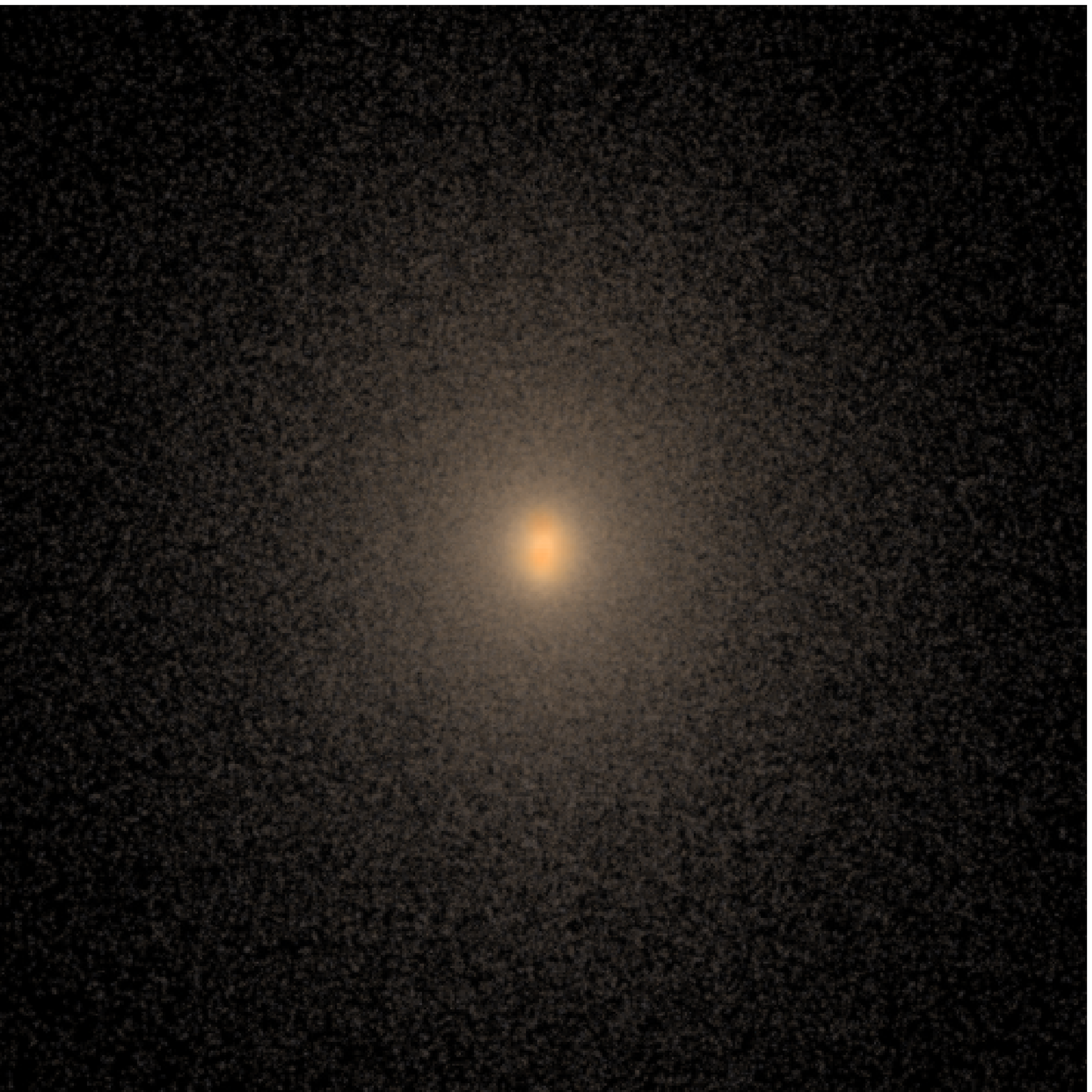,width=0.24\textwidth}
  \psfig{figure=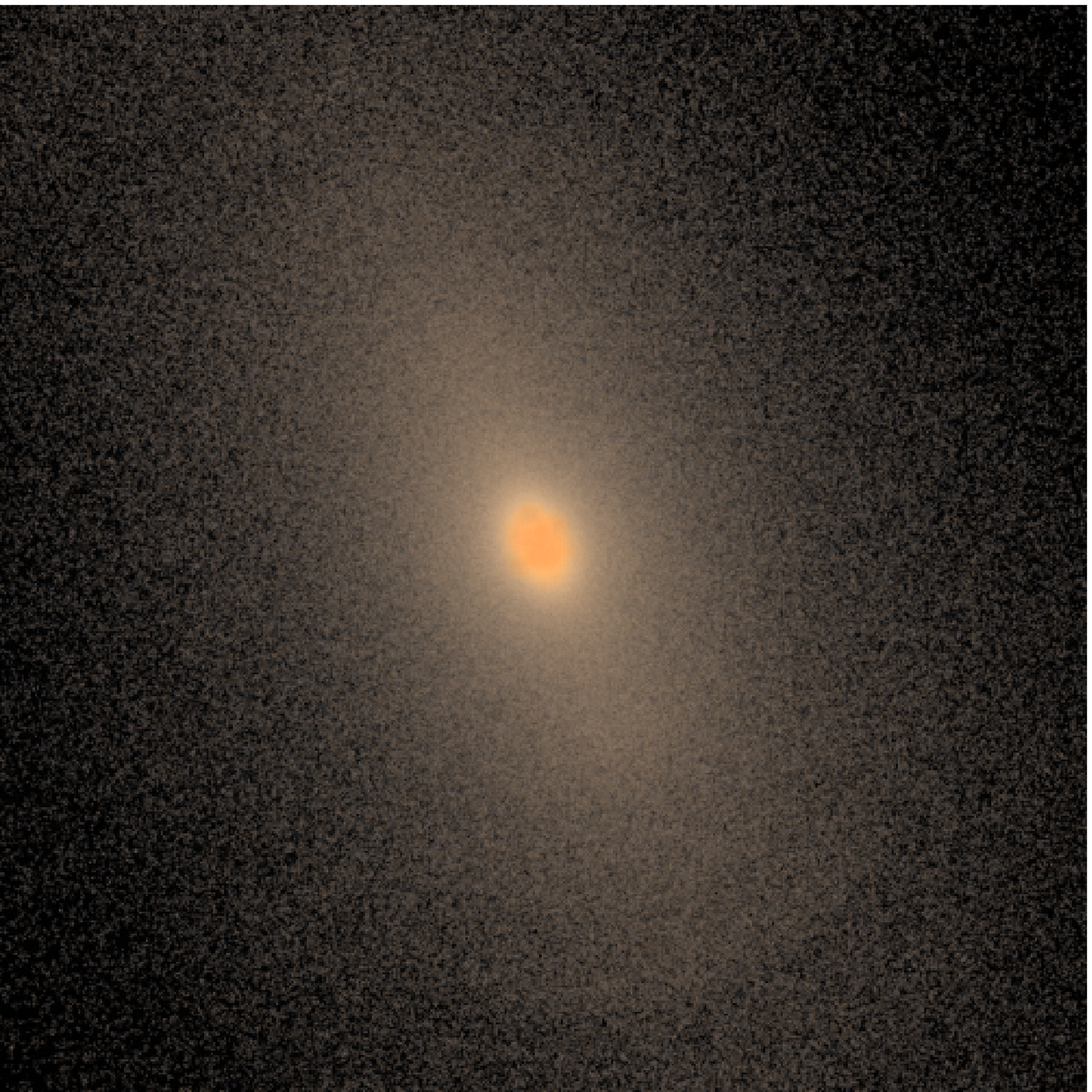,width=0.24\textwidth}
  \psfig{figure=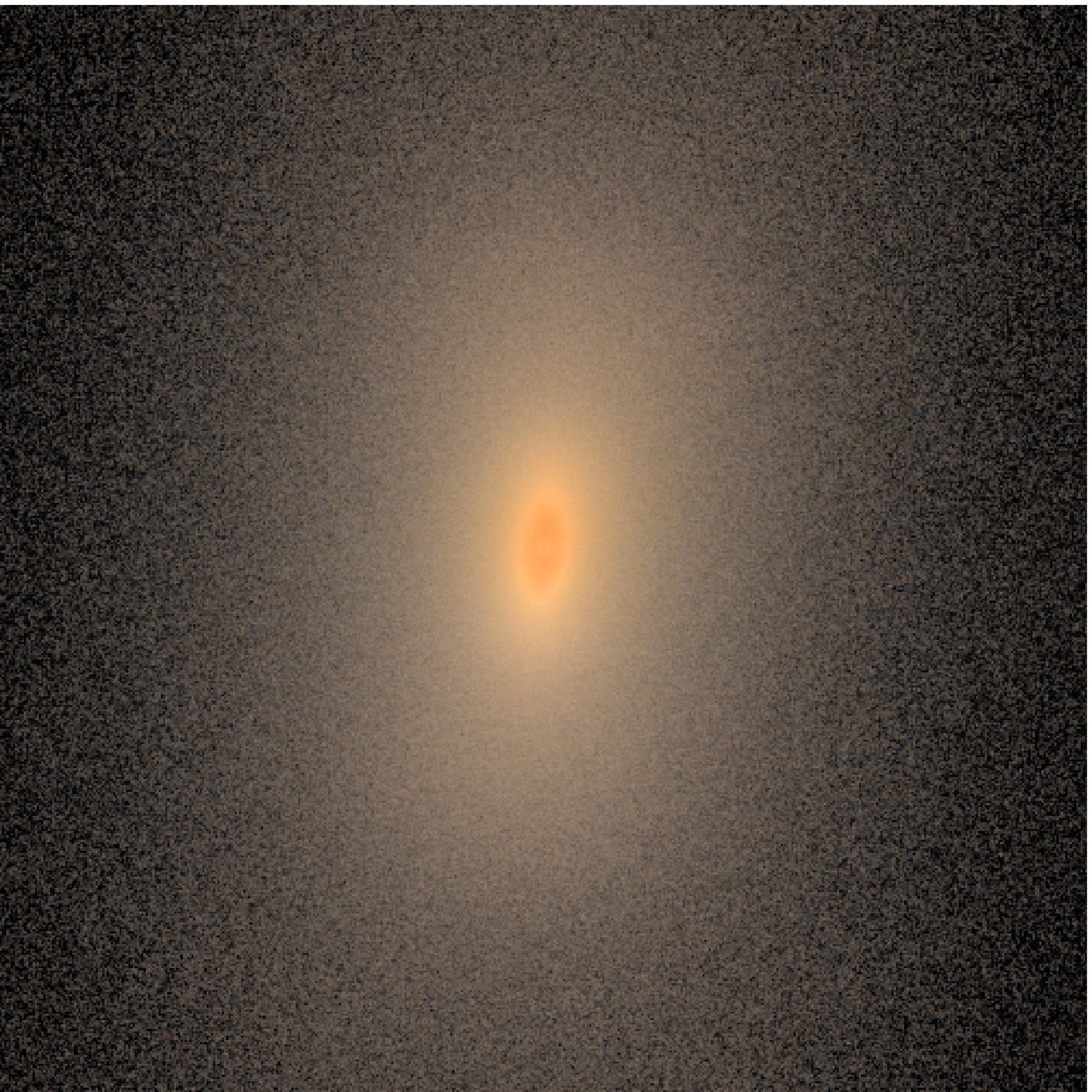,width=0.24\textwidth}
}
\caption{ {\sc{sunrise}} images of forced quenching simulations of
  halo4 at redshift $z=0$ with resolution increasing from left to
  right.}
  \label{fig:images2}
\end{figure*}

\subsection{Mass density slopes}
A further observational test of the realism of the simulated galaxies
comes from the slope of the total mass profile, $\gamma'$, which is
measured near the galaxy half-light radius.  Fig.~\ref{fig:gammap}
shows $\gamma'$ vs stellar surface density, $\Sigma_{\rm
  star}=\Mstar/(2\pi R_e^2)$ for our simulations and observations at
$z\sim 0$.  Points with error bars show observations using strong
lensing from the SLACS survey (Auger \etal 2010b), where the stellar
density is calculated assuming a Salpeter IMF.  In our simulations we
measure $\gamma'$ between 1\% and 2\% of the virial radius, as this
corresponds to the typical half-light sizes of quiescent galaxies
(Kravtsov 2013). The forced quenching simulations (red filled symbols)
have $1.8 \lta \gamma'\lta 2.1 $, while the standard simulations
(black open symbols) have $2.3\lta \gamma'\lta 2.4$. Observed galaxies
exist  with all such values, though the forced quenching simulations
have more typical values, and while the forced quenching simulations
fall within the observed distribution of $\gamma'$ and $\Sigma_{\rm
  star}$, the standard simulations fall well outside.

It is also possible to measure the mass  slope over a longer baseline
using the velocities at $r_{1/2}$ and $R_{200}$.  Observations using
weak gravitational lensing and satellite kinematics find that for
massive ellipticals $V_{1/2}\sim V_{200}$ (Dutton \etal 2010).  The
standard simulations have declining circular velocity profiles with
$V_{1/2}/V_{200}\sim 1.6$ to $2.0$.  The forced quenching simulations
have much flatter profiles with $V_{1/2}/V_{200}\sim 0.8$ to $1.0$ in
agreement with observations.

The slope of the mass profile both locally at $\sim r_{1/2}$ and
globally between $r_{1/2}$ and $R_{200}$ further support the
conclusion that the FQ simulations have realistic distributions of
stars and total matter at small radii, while the standard simulations have
too much stellar and total mass at small radii.

\subsection{Resolution effects on galaxy structure}
One of our initial conditions (halo4) has been run at four different
resolution levels: level0 is the lowest and level3 is the highest (see
Fig.~\ref{fig:massres} \& Table~\ref{tab:setup}).  In
Figs.~\ref{fig:rmvm}-\ref{fig:vv} the galaxy structural parameters for
these simulations are shown with larger circles for higher resolution.
There is clearly significant variation in the galaxy structural
parameters at different resolutions. Here we focus our discussion on
the forced quenching simulations at $z=0$.

The stellar mass increases with resolution by 0.39 dex from lowest to
highest. However, most of this difference (0.33 dex) is due to the
lowest resolution simulation, which has {\it only} $\sim 10^{5}$ dark
matter particles within the virial radius.  There is just 0.06 dex
difference between level1 and level3. Circular velocity at the
half-mass radius, $V_{1/2}$, shows a similar qualitative trend, with
higher velocities in higher resolution simulations, but with smaller
relative changes.  Between the highest three resolution simulations
the variation in $V_{1/2}$ is just 2\%.  Half-mass sizes vary by a
factor of $\sim 2$ with no clear trend with resolution. Stellar
surface densities show the largest variation with a factor of $\sim 8$
difference.  This variation in surface density is clearly visible in
the galaxy images shown in Fig.~\ref{fig:images2}.  If we remove the
lowest resolution simulation the variations in sizes and surface
densities are reduced to a factor of $\sim 1.6$ and $\sim 2.0$,
respectively.

In summary, for simulations with more than a million particles inside
the virial radius the stellar masses and circular velocities are well
converged while the half-mass sizes and average stellar surface
densities show significant variation.

\section{Dark Halo Response to Galaxy Formation}
\label{sec:haloresponse}

\begin{figure}
\centerline{
\psfig{figure=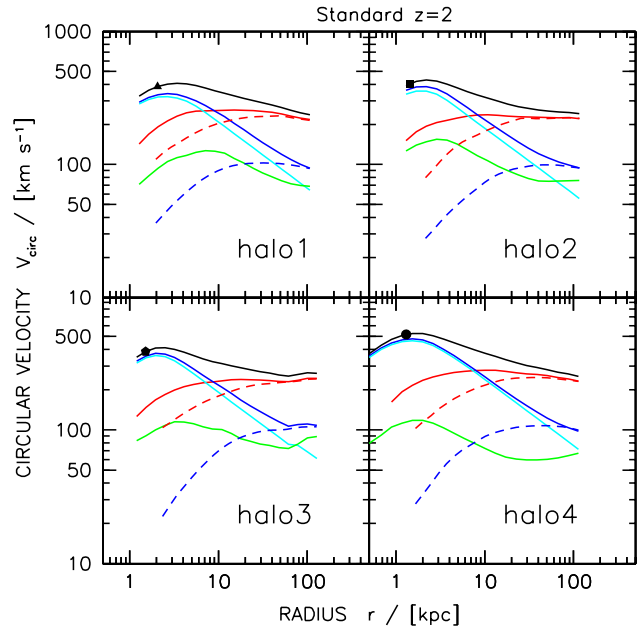,width=0.5\textwidth}
}
\caption{Circular velocity profiles for most massive galaxies in the
  standard simulations at $z=2.1$ (solid lines): Total (black); dark
  matter (red); baryons (blue); stars (cyan); gas (green). The dashed
  lines show the dark and baryons for the no cooling simulation, which
  acts as a control to the impact of galaxy formation. The filled
  symbol corresponds to the circular velocity at the stellar half-mass
  radius.}
\label{fig:vcirc_fid_z2}
\end{figure}

\begin{figure}
\centerline{
\psfig{figure=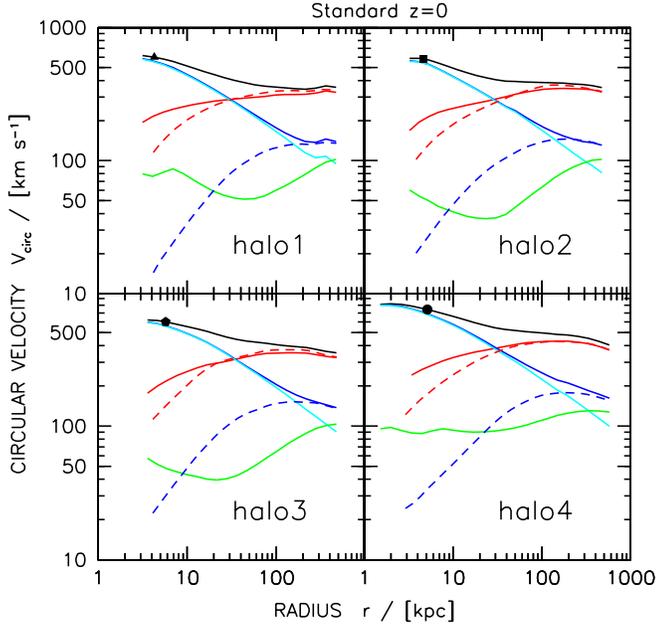,width=0.5\textwidth}
}
\caption{Circular velocity profiles for most massive galaxies in the
  standard simulations at $z=0$ (solid lines): Total (black); dark
  matter (red); baryons (blue); stars (cyan); gas (green). The dashed
  lines show the dark and baryons for the no cooling simulation, which
  acts as a control to the impact of galaxy formation. The filled
  symbol corresponds to the circular velocity at the stellar half-mass
  radius.}
\label{fig:vcirc_fid_z0}
\end{figure}

In the previous section we have established how well our various
simulations reproduce observed global properties of elliptical
galaxies at $z=0$ and their progenitors at $z\sim 2$.  We now turn our
attention to the radial distribution of dark matter, and in particular
how this responds to the galaxy formation process.

\subsection{Circular velocity profiles}
\begin{figure}
\centerline{
\psfig{figure=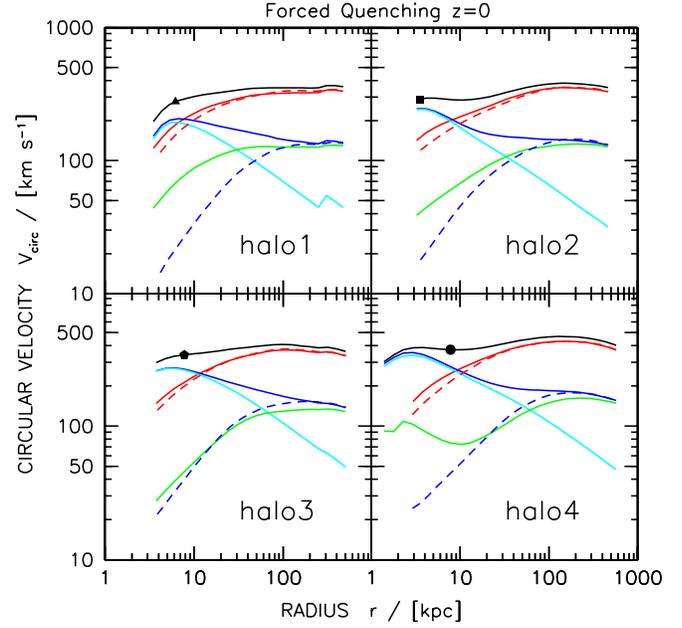,width=0.5\textwidth}
}
\caption{Circular velocity profiles for most massive galaxies in the
  forced quenching simulations at $z=0$ (solid lines): Total (black); dark
  matter (red); baryons (blue); stars (cyan); gas (green). The dashed
  lines show the dark and baryons for the no cooling simulation, which
  acts as a control to the impact of galaxy formation. The filled
  symbol corresponds to the circular velocity at the stellar half-mass
  radius.}
\label{fig:vcirc_hq_z0}
\end{figure}

Figs.~\ref{fig:vcirc_fid_z2}-\ref{fig:vcirc_hq_z0} show circular
velocity profiles, where $V_{\rm circ}=\sqrt{G M(r)/r}$, for our simulated
galaxies at $z=2.1$ and $z=0$. Velocity profiles are plotted from
$r_{\rm min}$ to the the virial radius, $R_{200}$. Here $r_{\rm min}$
is the maximum of the convergence radius as defined by Power \etal
(2003) and twice the softening length.  The circular velocity at
stellar half-mass radius is shown with a different symbol for each
halo, following on from previous plots.

When making these velocity profiles we have taken care to ensure that
the systems are relaxed.  During a merger event the assumption of
spherical symmetry breaks down, the center is no longer well defined,
and the derived mass profile will have an apparent core in the
center. By comparing the mass profiles at different snapshots spaced
nearby in time, the presence of mergers is easily detected.  We have
selected the snapshot closest to $z=0$ or $z=2.1$ in which both the
galaxy formation and no cooling simulations show no obvious signs of
being compromised by mergers.

\begin{figure}
\centerline{
\psfig{figure=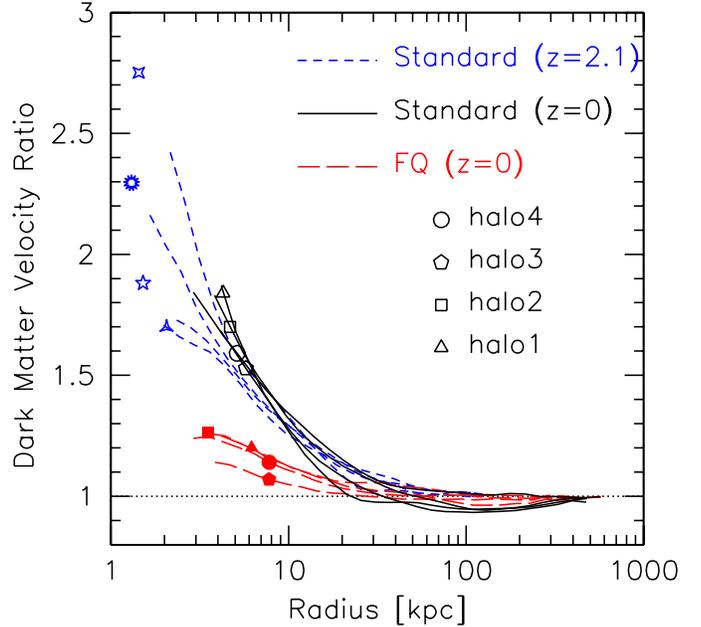,width=0.5\textwidth}  
}
\caption{Change in dark matter circular velocity due to galaxy
  formation: with standard physics at $z=2.1$ (blue, short-dashed
  lines) at $z=0$ (black, solid lines); and with forced quenching at
  $z=0$ (FQ, red, long-dashed lines). The filled symbols indicate the
  half-mass radii of the stars.  All simulations contract within $\sim
  20$ kpc, but there is substantially less contraction in the FQ
  simulations.}
\label{fig:vratio}
\end{figure}

\begin{figure*}
\centerline{
\psfig{figure=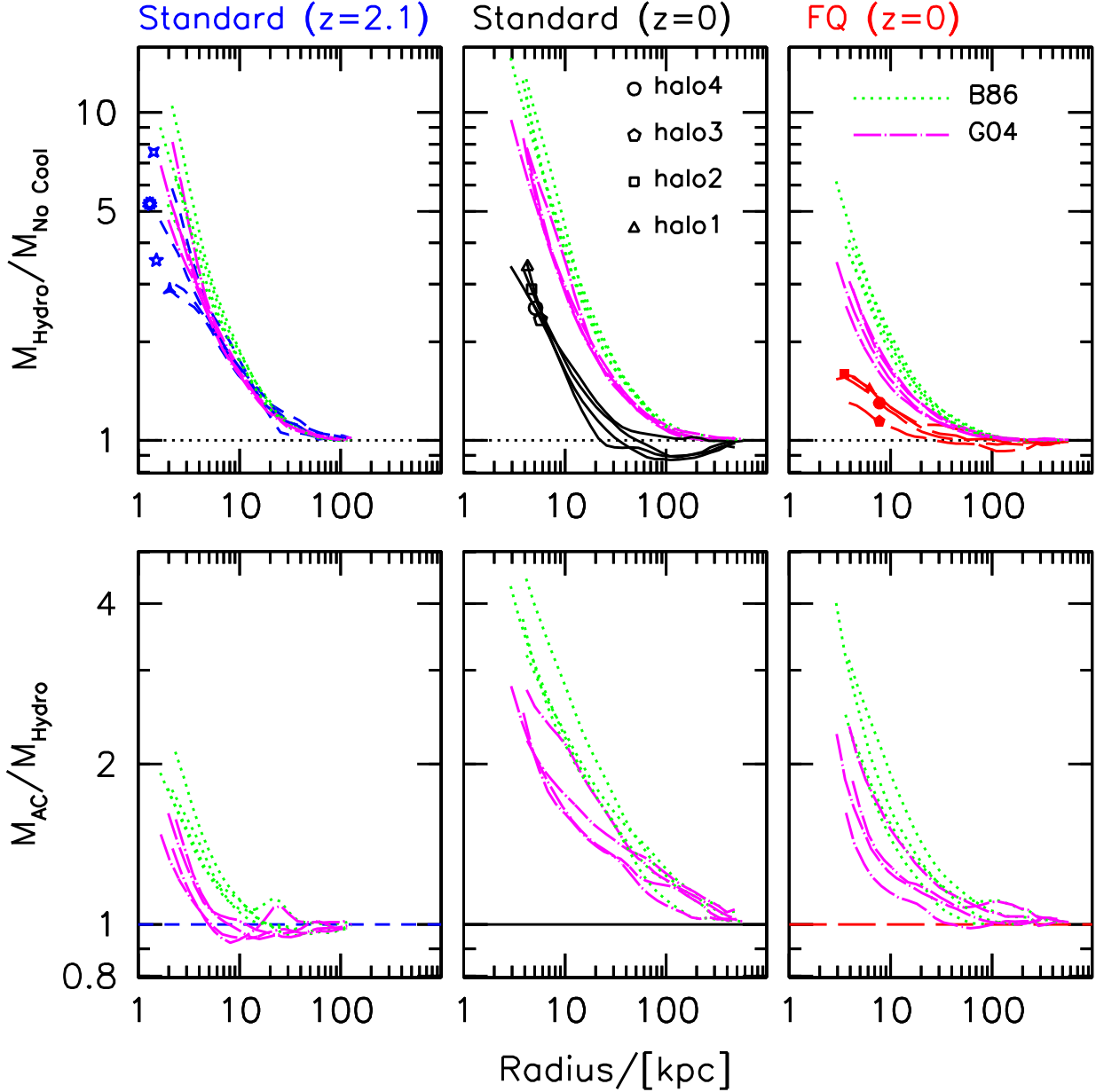,width=0.95\textwidth}  
}
\caption{Change in enclosed dark matter mass due to galaxy formation:
  with standard physics at $z=2.1$ (left) at $z=0$ (middle); and with
  forced quenching at  $z=0$ (right).   Predictions from the adiabatic
  contraction (``AC'') models of Blumenthal \etal (1986, B86) and Gnedin \etal
  (2004, G04) are shown with green dotted and magenta dot-dashed
  lines, respectively. At redshift $z=0$ the contraction in our
  simulations is substantially weaker than predicted by these models.
  Upper panels show change with respect to simulations without cooling
  (``No Cool''), while lower panels show differences with respect to
  the galaxy formation simulations (``Hydro'').}
\label{fig:vratio2}
\end{figure*}

A comparison of the circular velocities at the virial radius (the last
point in the velocity profile) shows that the standard (solid lines,
Fig.~\ref{fig:vcirc_fid_z0}), forced quenching (solid lines,
Fig.~\ref{fig:vcirc_hq_z0}) and no cooling (dashed lines,
Figs.~\ref{fig:vcirc_fid_z0} \& \ref{fig:vcirc_hq_z0}) simulations all
have the same dark and baryonic masses. Furthermore, this implies that
the haloes have been able to retain all of their cosmic share of
baryons.  In all simulations stars (cyan lines) dominate the baryon
budget at small radii, while gas (green lines) becomes as important as
the stars at large radii.  In the FQ simulations the gas roughly
follows the mass profile of the dark matter, but in the standard
simulations there is an additional component of concentrated cold gas.

A comparison between the solid and dashed red lines in
Figs.~\ref{fig:vcirc_fid_z2} - \ref{fig:vcirc_hq_z0} shows that all the
dark matter haloes have contracted in response to galaxy
formation. The strength of the contraction is more clearly seen in
Fig.~\ref{fig:vratio} which shows the ratio of dark matter circular
velocities.  The numerator is the dark matter circular velocity in the
simulations with galaxy formation (i.e., standard/FQ) and the
denominator is the dark matter circular velocity in the simulations
without galaxy formation (i.e., no cooling).  In the standard
simulations the velocity ratio is roughly the same at $z=0$ (black
lines) as it is at $z=2.1$ (blue lines), whereas in the FQ simulations
(red lines) the contraction is not as strong.

\subsection{Adiabatic contraction formalism}

To understand the effect of the baryonic physics on the dark matter
density profile, we employ the analytic adiabatic contraction
formalism outlined in Blumenthal \etal (1986), and introduced in
another context in Barnes \& White (1984).  The main assumption is
that the time scale for galaxy formation is long compared to the
orbital time scale of the dark matter particles. With the simplifying
assumptions of spherical symmetry and circular dark matter particle
orbits, the  adiabatic invariant reduces to $r M(r) =const$, where
$M(r)$ is the total mass enclosed within radius $r$.

Thus given a spherically enclosed mass profile from a simulation without
gas cooling, $\Mi(r)$, (where the i refers to initial), we can 
derive the final dark matter profile, $M_{\rm dm,f}(r)$, once the
final baryonic mass profile, $M_{\rm bar,f}(r)$, is specified. The
initial total mass profile is split into dark matter $M_{\rm
  dm,i}(r)$, and baryons $M_{\rm bar,i}(r)$ either explicitly (as in our
no cooling simulations) or implicitly assuming $M_{\rm bar,i}(r) =
(1-\fbar)\Mi(r)$, where $\fbar$ is the cosmic baryon fraction
($\Omegab/\Omegam \sim 0.16$).

An additional  assumption is that the dark matter shells do not cross: $M_{\rm
  dm,f}(\rf)=M_{\rm dm,i}(\ri)$, where \ri is the ``initial'' radius
of a shell of dark matter, and \rf is the radius of this shell after
the effects of galaxy formation are included. Putting this together yields
\begin{equation}
\label{eq:ac}
\rf/\ri = \Mi(\ri)/[M_{\rm b,f}(\rf)+M_{\rm d,i}(\ri)].
\end{equation}
Thus given $\Mi$ and $M_{\rm b,f}$, one can solve Eq.~\ref{eq:ac} for
the mapping between $\rf$ and $\ri$, and hence derive the final dark
matter profile.  {\it The appeal of this formalism is that in the
adiabatic limit the response of the halo depends only on the final
state of the baryons, and is independent of how it was assembled.}

Particle orbits in CDM haloes are not circular. To account for this
Gnedin \etal (2004) introduced a modified adiabatic invariant: $r
M(\bar{r})$, where $r$ and $\bar{r}̄$ are the current and
orbit-averaged particle positions. The orbit average radius can be
approximated as: $\bar{r}\approx \Rvir A (r/\Rvir)^w$, with $A\approx
0.85$ and $w\approx 0.8$. This modified adiabatic invariant results in
slightly weaker contraction.

Fig.~\ref{fig:vratio2} shows the change in the dark matter masses
profiles in our simulations and that predicted by  the adiabatic
contraction models of Blumenthal \etal (1986, B86) and Gnedin \etal
(2004, G04). At $z=2.1$ (left panel) the G04 model works for radii $r
\gta 5$ kpc, and the B86 model for $r \gta 10$ kpc. At smaller radii
the models predict more contraction than seen in our simulations. By
$z=0$ (middle and right panel) the over prediction has grown and
extends to all radii (less then the virial radius). For example, at 5
kpc the G04 and B86 models over predict the mass by a factor of $\sim
2$ and $\sim 3$, respectively.  These results are qualitatively
similar to previous studies on Milky Way mass haloes (Abadi \etal
2010; Pedrosa \etal 2010).

\begin{figure}
\centerline{
 \psfig{figure=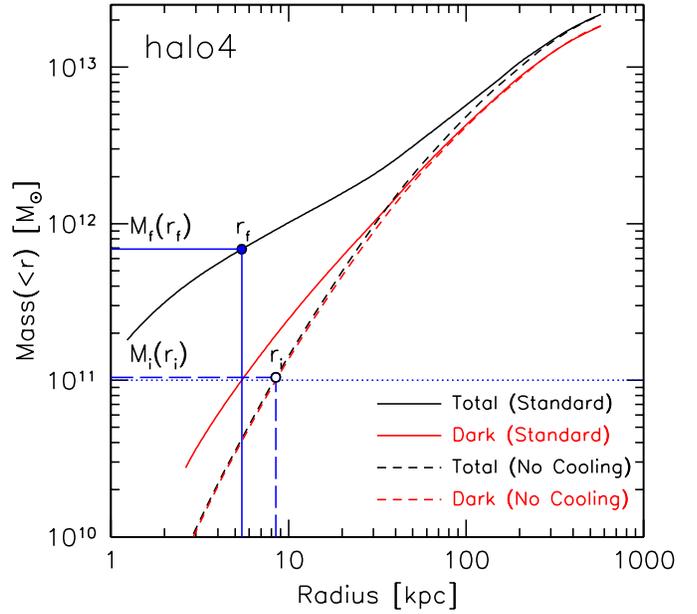,width=0.5\textwidth} 
}
\caption{Example showing how $\rf, \ri, \Mf, \Mi$ are calculated in
  the cumulative mass vs radius plot for simulation Halo4 using
  standard galaxy formation physics.  The horizontal blue dotted line
  shows an arbitrary mass, here $10^{11}\Msun$. The radius this
  intersects the dark matter profile from the standard simulation is
  termed $\rf$ (vertical blue solid line), while the radius this
  intersects the dark matter profile from the no cooling simulation
  simulation is termed $\ri$ (vertical blue dashed line). The total
  masses contained within $\rf$ and $\ri$ are then given by $\Mf$
  (horizontal blue solid line) and $\Mi$ (horizontal blue dashed
  line) respectively.}
\label{fig:rfi}
\end{figure}

Returning to the halo response formalism, an example of how the
radii and masses are calculated in one of our simulations is shown in
Fig.~\ref{fig:rfi}. The solid lines show the total (black) and dark
matter (red) mass profiles for the standard simulation for
halo4.2. The dashed lines show the total (black) and dark matter (red)
mass profiles for the corresponding simulations with no cooling.  The
horizontal dotted blue line shows an arbitrary mass (here
$10^{11}\Msun$), which intersects the final dark matter mass profile
at $\rf$ and the initial dark matter mass profile at $\ri$. \Mf and
\Mi are then the total enclosed mass at \rf and \ri, respectively.
This process is repeated for different arbitrary masses to obtain the
relation between $\rf/\ri$ and $\Mi/\Mf$.

\begin{figure*}
\centerline{
  \psfig{figure=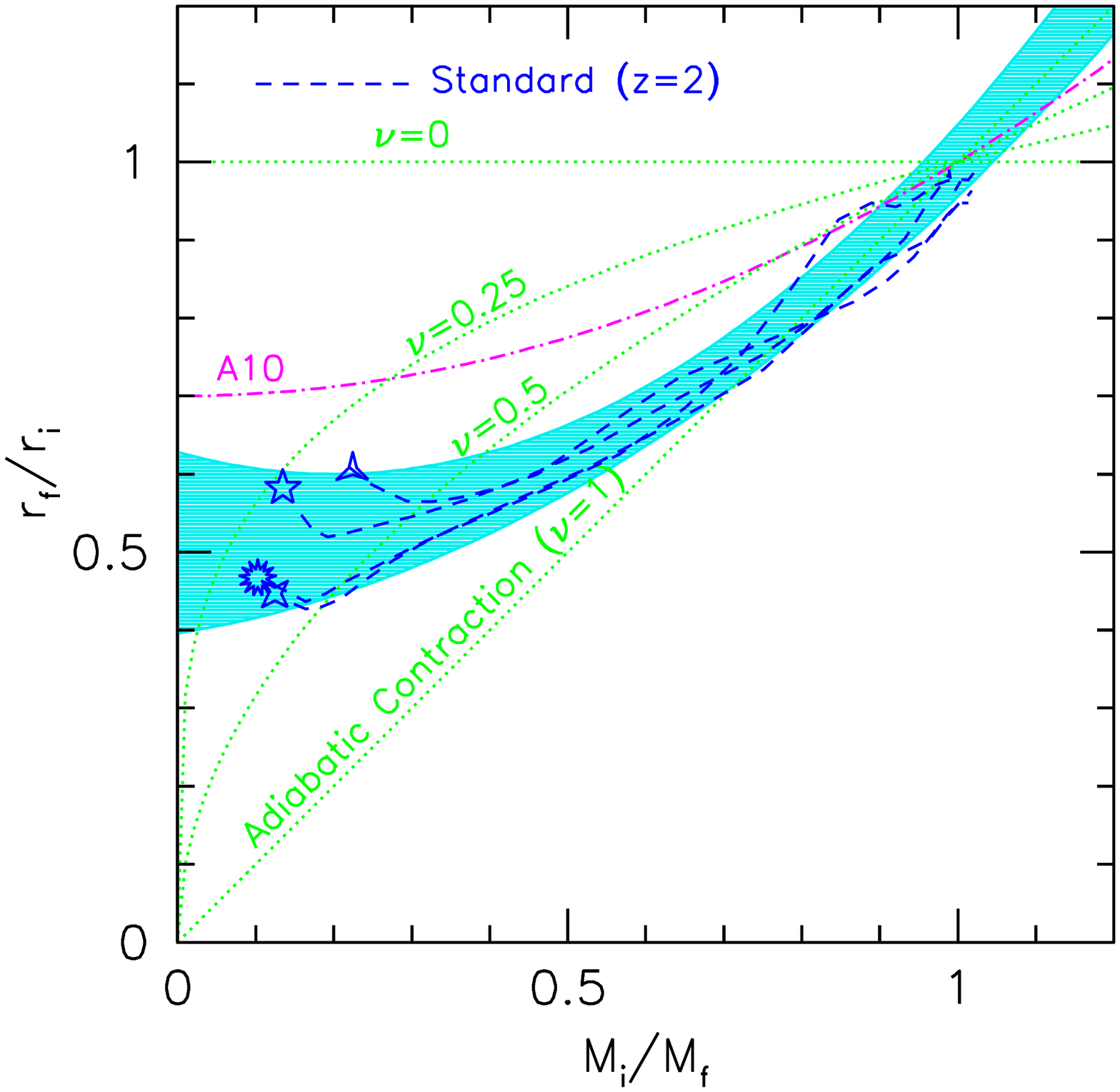,width=0.45\textwidth}  
  \psfig{figure=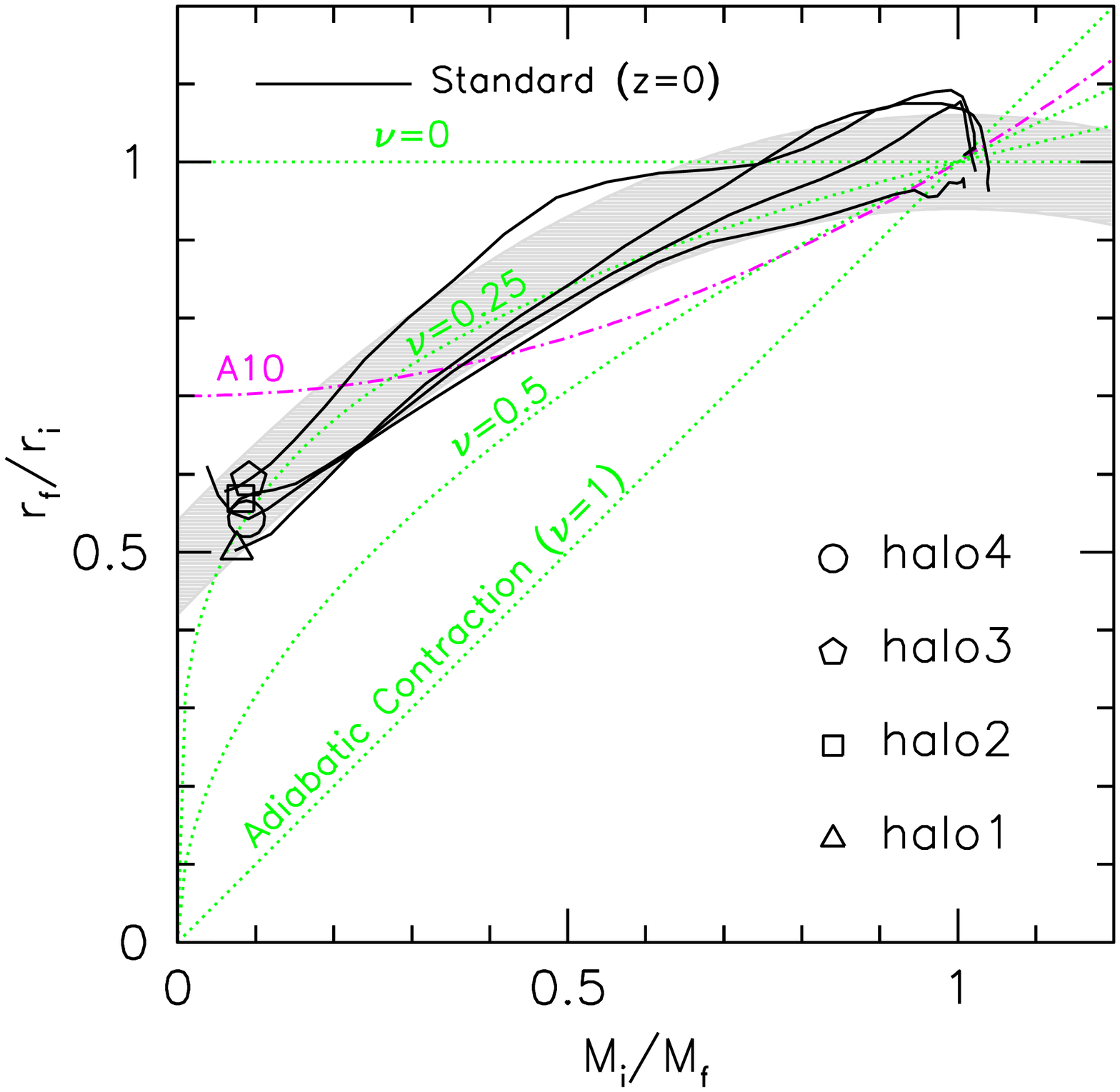,width=0.45\textwidth}  
}
\centerline{
  \psfig{figure=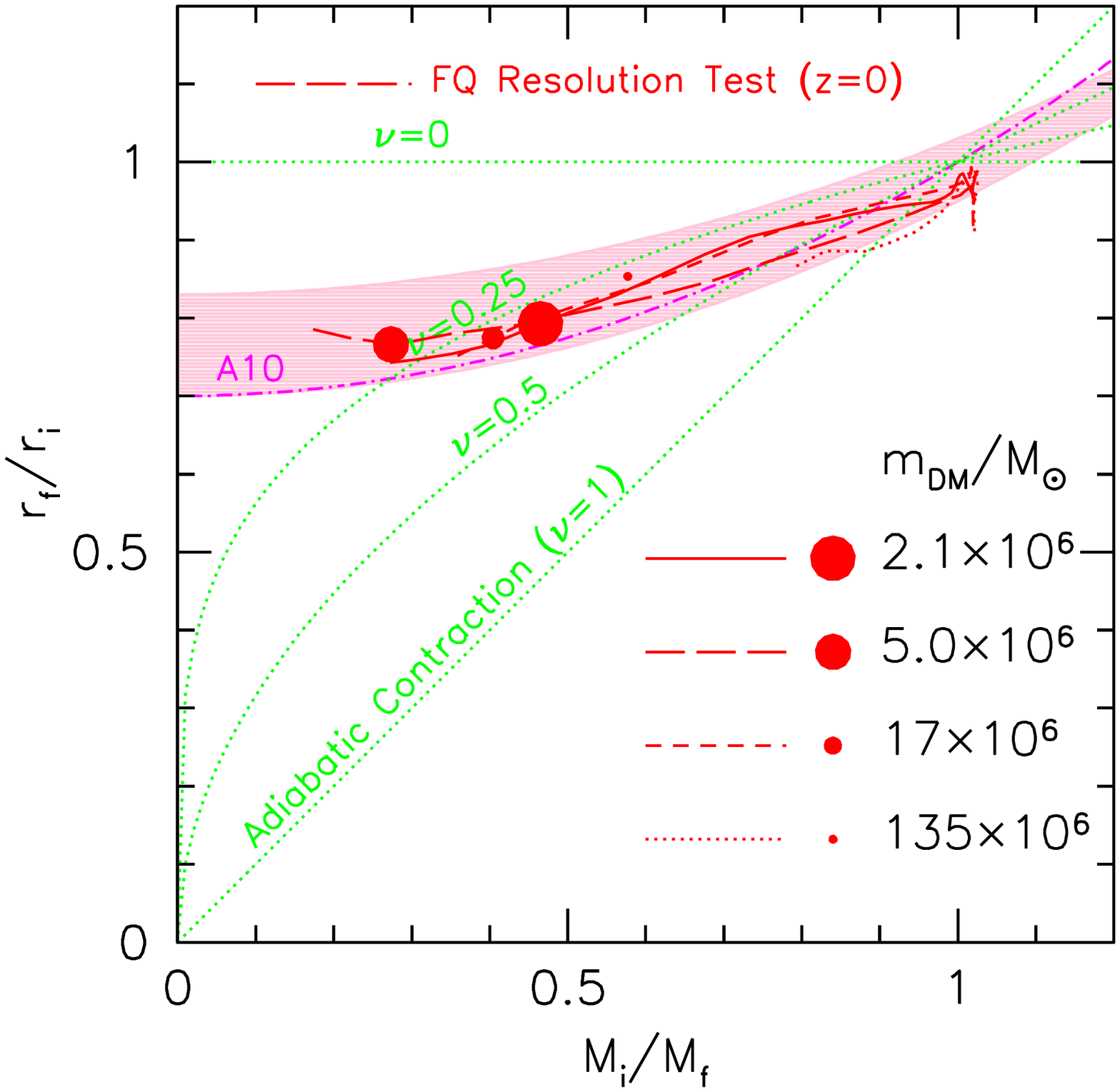,width=0.45\textwidth}  
  \psfig{figure=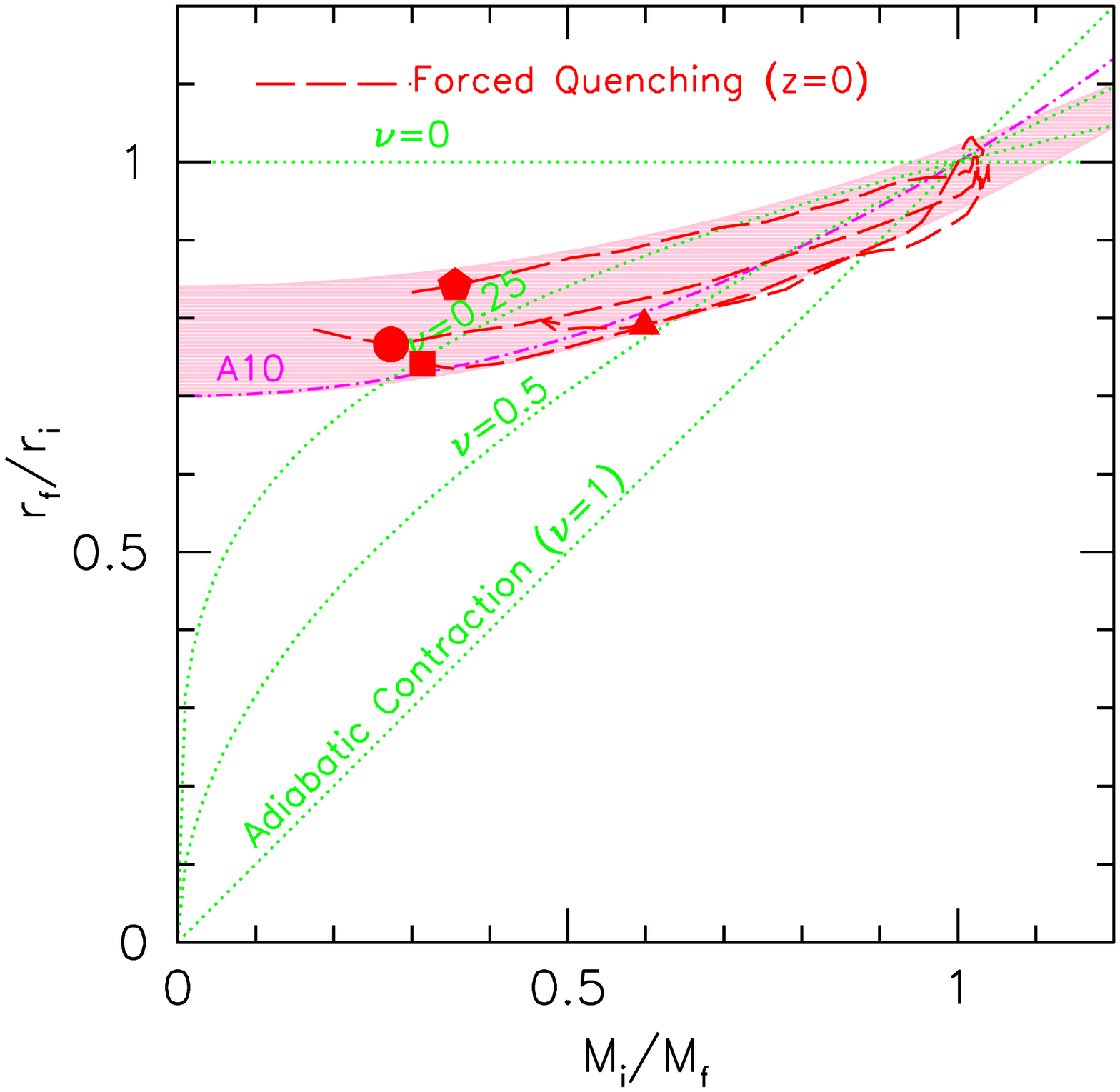,width=0.45\textwidth}  
}
\caption{Dark halo response of our simulations: standard $z=2.1$ (upper
  left, blue); standard $z=0$ (upper right, black); forced quenching
  $z=0$ (lower right, red); forced quenching $z=0$ resolution tests
  with halo4 (lower left, red).  The filled points show the halo
  response at the galaxy half-stellar mass radius (with different
  simulations represented by different symbols as indicated). The
  dotted green lines show various halo responses of the form
  $\rf/\ri=(\Mi/\Mf)^{\nu}$. Where $\nu=1$ corresponds to adiabatic
  contraction, $\nu=0$ corresponds to no change, and $n=0.5$ and
  $n=0.25$ correspond to weak contraction. The dot-dashed magenta
  lines show the halo response model of Abadi \etal (2010, A10).}
\label{fig:ac}
\end{figure*}

Fig.~\ref{fig:ac} shows the tracks of our simulations in the $\rf/\ri$
vs $\Mi/\Mf$ plane. The y-axis shows the ``contraction'' factor, while
the x-axis can (typically) be mapped monotonically to radius. The
adiabatic contraction formula (Eq.~\ref{eq:ac}) predicts that
$\rf/\ri=\Mi/\Mf$, which is indicated by the diagonal dotted line in
the figure. No change in the dark matter profile corresponds to the
horizontal dotted line at $\rf/\ri=1$, and expansion to $\rf/\ri > 1$.

All simulations show contraction at small radii (see also
Figs.~\ref{fig:vcirc_fid_z2}-\ref{fig:vcirc_hq_z0}), with the standard
simulations at $z=0$ showing a small amount of expansion at large
radii, $r\gta 30$ kpc, which corresponds to the radius where there is
equal baryons and dark matter.
The points show the halo
response at $\rf=r_{1/2}$. The standard simulations at $z=2.1$ and $z=0$
have $\rf/\ri\sim0.5$, vs $\sim 0.1$ for adiabatic contraction, and
the FQ simulations have $\rf/\ri\sim0.8$ vs $\sim 0.4$ for adiabatic
contraction. 

Overall the halo response does not follow a single track, although
interestingly at a given redshift each type of simulation results in a
similar halo response.
To quantify different halo responses we first consider the equation
introduced by Dutton \etal (2007):
\begin{equation}
\label{eq:acnu}
\rf/\ri = (\Mi/\Mf)^{\nu}.
\end{equation}
Adiabatic contraction is thus $\nu=1$, contraction $0 < \nu < 1$, no
change to $\nu=0$, and expansion $\nu < 0$. The cases $\nu=0,0.25,0.5$
and $1$ are shown in Fig.~\ref{fig:ac} with dotted green lines.  We
note that the model of Gnedin \etal (2004) can be approximated with
$\nu \sim 0.8$, and thus can be considered moderately strong
contraction.

While the standard simulations at $z=0$ are approximated by
Eq.~\ref{eq:acnu} with $\nu=0.25$, the standard simulations at $z=2.1$
and the FQ simulations at $z=0$ are not well described by this
formula.  The magenta lines show the formula from Abadi \etal (2010),
$\rf/\ri = 1 -0.3(\Mi/\Mf -1)^2$, which approximately describes our
forced quenching simulations at $z=0$ (lower right panel). This
similarity may be a coincidence as their simulations are missing
important aspects of galaxy formation processes such as star formation
and feedback.

The shaded regions in Fig.~\ref{fig:ac} bracket
the halo response shown in the 4 initial conditions.
They are given by the following relations,
where the uncertainties bracket the range in halo response:
standard simulations at $z=2.1$ (upper left) 
\begin{equation}
\label{eq:s0}
  \rf/\ri = 0.5(^{+0.10}_{-0.12}) + 0.5(^{+0.2}_{-0.1})(\Mi/\Mf +0.0^{-0.2}_{+0.2})^2;
\end{equation}
standard simulations at $z=0$ (upper right) 
\begin{equation}
\label{eq:s2}
\rf/\ri = 1.0(^{+0.06}_{-0.06}) -0.52(\Mi/\Mf -1)^2;
\end{equation}
FQ simulations (lower right)
\begin{equation}
\label{eq:fq0}
\rf/\ri = 0.75(^{+0.07}_{-0.07}) + 0.25(^{-0.03}_{+0.03})(\Mi/\Mf)^2;
\end{equation}

A numerical convergence test is shown in the lower left panel of
Fig.~\ref{fig:ac}.  This shows the halo response for halo4 run at four
different mass (and force) resolution levels (see Table
~\ref{tab:setup}). All simulations have similar halo response
(especially the highest three resolution simulations), suggesting
convergence. The lowest resolution simulation (halo4.0) does not have
as much contraction, but follows a similar path in the $\rf/\ri$ vs
$\Mi/\Mf$ plane.

\begin{figure}
\centerline{
\psfig{figure=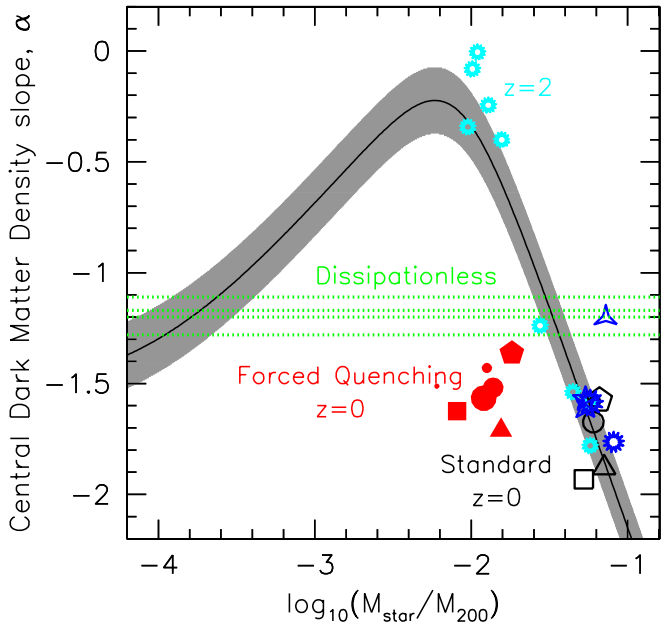,width=0.5\textwidth}
}
\caption{Logarithmic slope of the dark matter density profile,
  $\alpha$,  measured between 1 and 2\% of the virial radius,
  $R_{200}$, vs the ratio between the stellar and virial mass, $M_{\rm
    star}/M_{200}$. Blue symbols show standard simulations of halo4 at
  $z=2.1$, black symbols show standard simulations at $z=0$, and red
  symbols show forced quenching simulations at $z=0$. The green dotted
  lines show $\alpha\sim -1.2$ for the control simulations (without
  cooling). The grey shaded region shows the relation  from Tollet
  \etal (2015, in prep) which was calibrated against haloes of mass
  $10^{10} - 10^{12} \Msun$.}
\label{fig:dicintio}
\end{figure}

\subsection{Central dark matter slopes}
Another way to express the strength of the contraction of the dark
matter halo is the central density slope, $\alpha$. We measure
$\alpha$ between $0.01$ and $0.02R_{200}$ which corresponds to $\sim 5
- 10$ kpc for our simulations at $z=0$.  Following Di Cintio \etal
(2014) we show in Fig.~\ref{fig:dicintio} the relation between
$\alpha$ and galaxy formation efficiency, $\Mstar/M_{200}$.  Our FQ
simulations (red points) have $-1.7 \lta \alpha \lta -1.4$, while the
standard simulations (black points) have $-2.0 \lta \alpha \lta -1.6$.
For reference, the dotted green lines show the no cooling simulations
which have $-1.3 \lta \alpha \lta -1.1$, the same values as found in
dark matter only simulations.

The solid line and shaded region shows the relation obtained from
Tollet \etal (2015, in prep) who used a set of $\sim 80$ zoom-in
simulations of halo mass $\sim 10^{10}$ to $\sim 10^{12}$ from the
NIHAO project (Wang \etal 2015). Our standard simulations are close to
this relation (both at $z=0$ and $z=2.1$), but the forced quenching
simulations are significantly offset. At a stellar to halo mass ratio
of $\Mstar/M_{200}\sim 0.01$ spiral galaxy haloes have large amounts
of expansion, whereas our elliptical simulations contract. Thus the
inner dark matter density slope is not purely determined by the
efficiency of star formation.  This difference is a reflection of the
different formation channels of spiral vs elliptical galaxies. In
spiral galaxies the vast majority of stars form in situ, with the
efficiency being regulated by SN feedback which also has strong
expansive effects on the halo.  In elliptical galaxies, a significant
fraction of the stars were formed ex situ, and were assembled with dry
mergers which has a milder expansive effect on the halo.

Finally, we note that the form of the halo response in
Eqs.~\ref{eq:s0}-\ref{eq:fq0}, which has a constant ratio between
initial and final radii at $r=0$, implies that at very small radii the
central dark matter slope tends to the dissipationless case. If haloes
are described by the Navarro, Frenk \& White (1997) formula, then the
inner slope would be $-1$, however, if the Einasto (1965) profile
holds then this implies the centers of dark matter haloes have
constant density.  Practically speaking, however, the presence of a
supermassive black hole may cause a dark matter spike (Gondolo \& Silk
1999), rendering these extrapolations a moot point.

\begin{figure*}
\centerline{
\psfig{figure=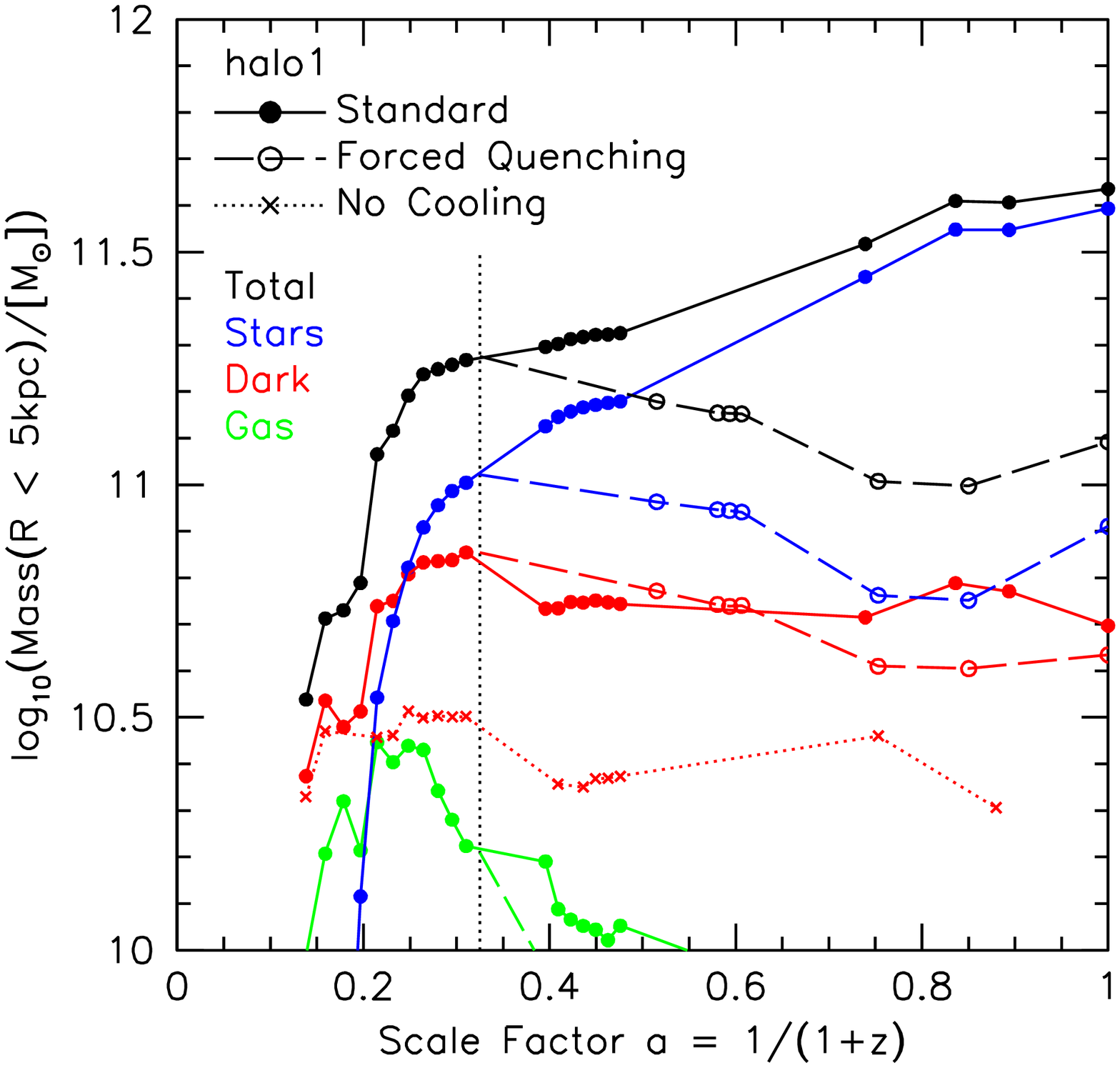,width=0.45\textwidth}
\psfig{figure=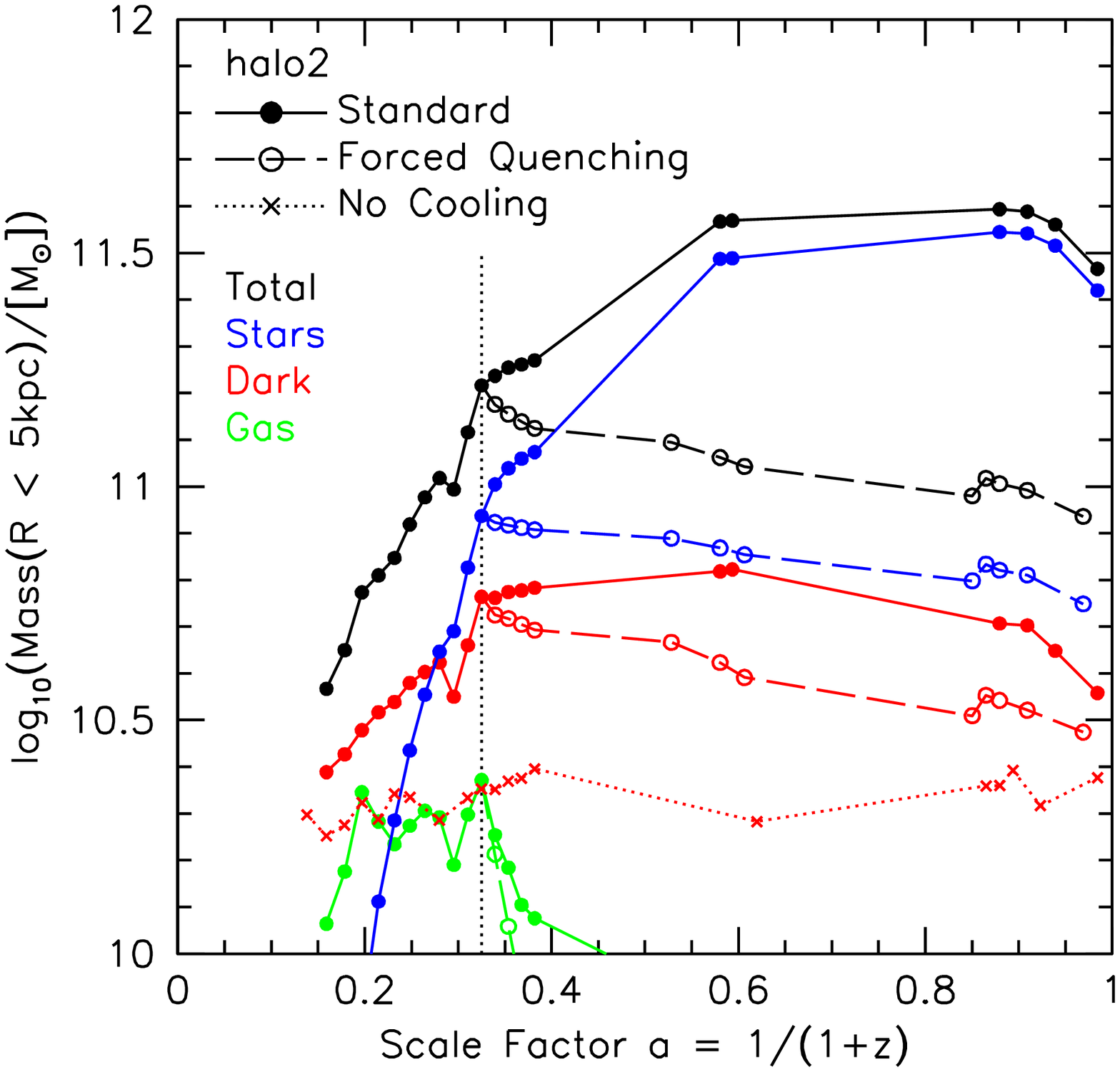,width=0.45\textwidth}
}  
\centerline{
\psfig{figure=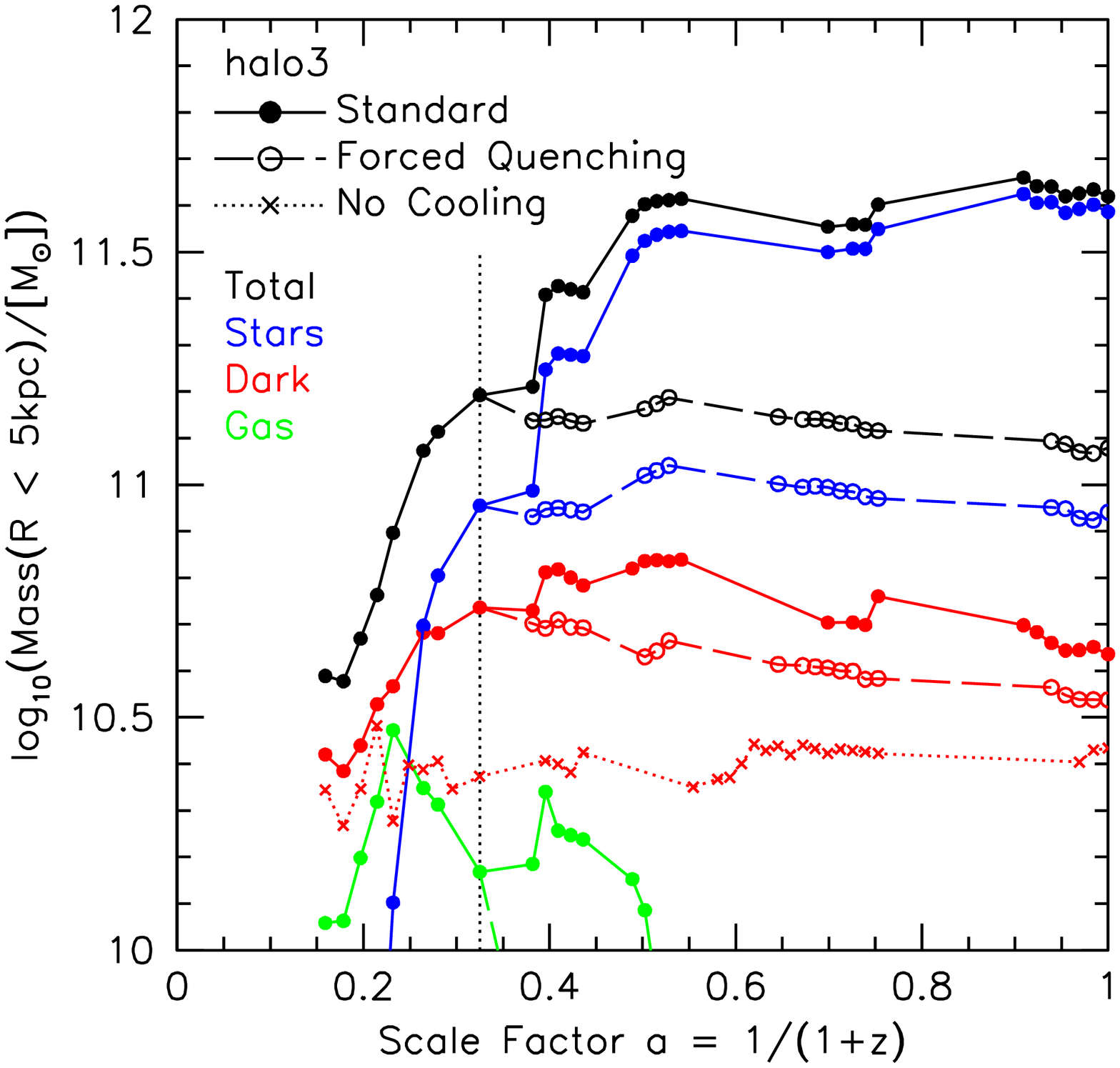,width=0.45\textwidth}
\psfig{figure=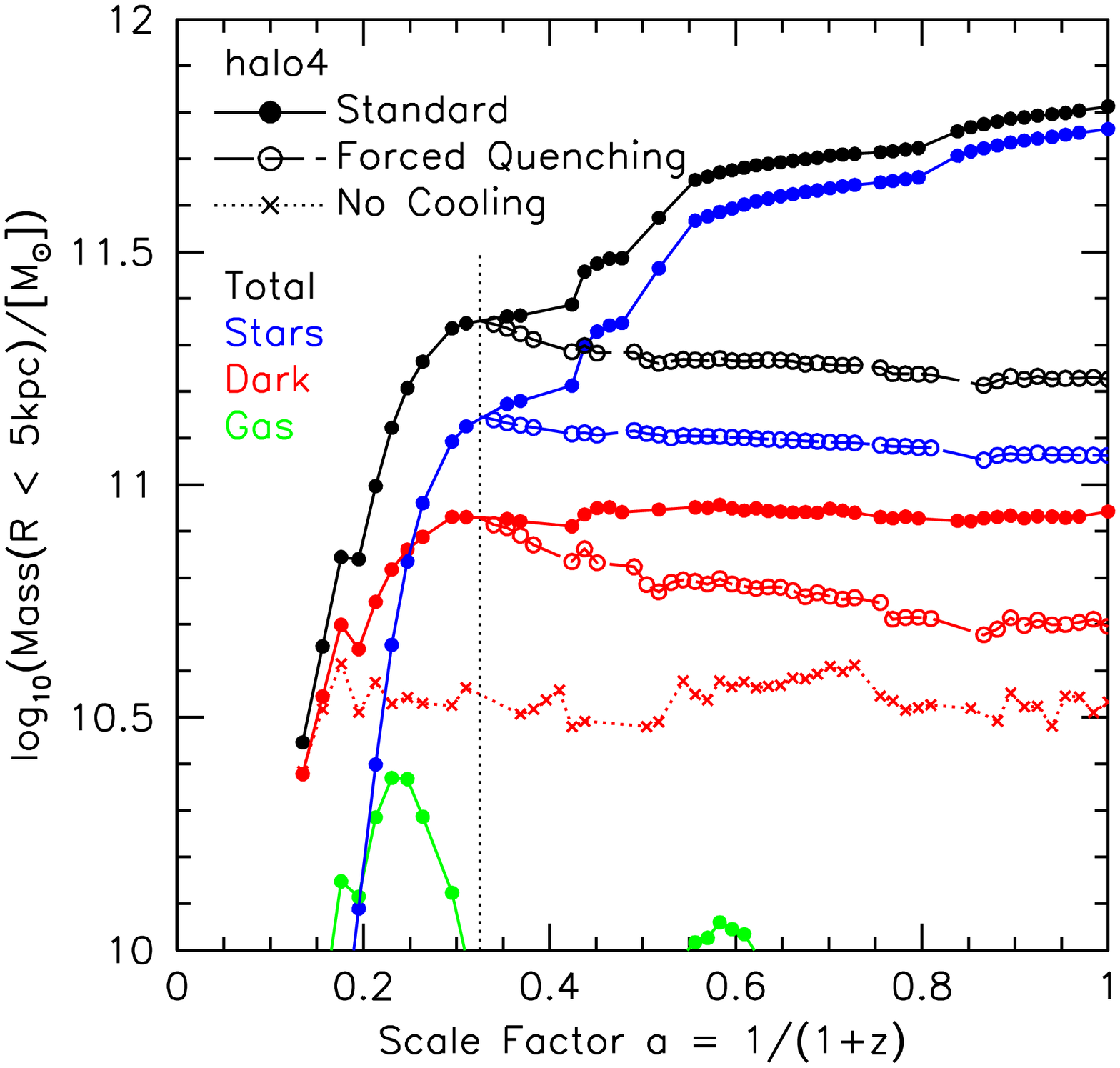,width=0.45\textwidth}
}
\caption{Mass enclosed within 5 proper kpc vs scale factor for the
  most massive progenitors in our 4 simulations. The total mass is
  given by black lines and points, dark matter by red and stars by
  blue. In the no cooling simulations (crosses, dotted line) the dark
  matter is in place by $a\sim 0.15$  $(z\sim 5)$. In the standard
  simulations (filled circles, solid lines) the halo contracts,
  increasing the mass by a factor $\sim 2.5$. The contraction occurs
  by $z=2.1$ and remains constant to $z\sim 0$, even though the
  stellar mass increases by an order of magnitude during thus
  interval. In the forced quenching simulations (open circles, dashed
  lines) the dark matter within 5 kpc gradually decreases with time
  due to a combination of stellar mass loss mass and dry minor
  merging.}
\label{fig:dm_z}
\end{figure*}

\subsection{Slow expansion}

When we disable gas cooling at $z=2.1$, the gas at small radii heats up
and expands to larger radii, see the green lines in
Fig.~\ref{fig:dm_z}.  One might worry that this would cause the halo
to expand artificially. However, the circular velocity plots in
Fig.~\ref{fig:vcirc_fid_z2} show that the gas is a sub-dominant
component with $V_{\rm gas}\sim 0.1 V_{\rm circ}$, and thus
contributes just $\sim 1\%$ to the enclosed mass at all radii (recall
that $M\propto V^2$).  Thus in the extreme case that all of
the gas is removed instantaneously, the expansion of the dark and
stellar mass distributions would be negligible.  

Fig.~\ref{fig:dm_z} shows the evolution in the mass enclosed within 5
proper kpc for our simulations. The total mass is shown in black, the
dark matter in red, the stars in blue and the gas in green.
The three line types correspond to the standard (solid), FQ (long-dashed),
no cooling (dotted) simulations. Time steps where the halo is clearly
undergoing a merger have been removed since it is impossible to
determine the central 5 kpc accurately during those time periods.

In the no cooling simulation the dark matter mass is roughly 
constant at $\sim 2-3\times 10^{10}\Msun$ since $a=0.15$ $(z\sim 5)$.
In the standard simulation the dark halo starts to contract at $z\sim
5$, reaches a maximum mass a factor of $\sim 2.5$ times that of the no
cooling simulation at $z\sim 2$, and remains roughly constant to
$z=0$. By contrast the stellar mass within 5 kpc continues to increase
all the way to $z=0$.
In the FQ simulation the stellar mass is roughly constant since $z=2.1$,
with a slight reduction due to stellar mass loss, while the dark
matter mass slowly decreases since $z\sim2$ by $\sim 0.2$ dex.  This
gradual expansion of the dark matter halo since $z=2.1$ in the forced
quenching simulations is further evidence that the expansion is not
caused by a  single event at $z=2.1$ or later, such as a major merger.

There are two processes that are likely contributing to the smooth
expansion: multiple minor mergers (e.g., El-Zant \etal 2001) and
adiabatic expansion due to stellar mass loss. In our simulations up to
40\% of the stellar mass, i.e., $\sim 4\times 10^{10}\Msun$,  is
returned to the ISM. In the standard simulations this gas gets turned
into new stars, but in our forced quenching simulations this gas gets
heated and expands to larger radii resulting in adiabatic
expansion. This stellar recycling explains why the stellar masses of
our forced quenching simulations declines gradually with time, except
for the occasionally merger. The signature of the adiabatic expansion
processes is that within a fixed radius the relative reduction in dark
matter should equal the relative reduction in stellar mass. This
appears to be the case for halo1 and halo2, but halo3 has a merger
event at $a\sim0.45$ which increases the stellar mass, but decreases
the dark mass, and halo4 has stronger reduction in dark mass than
stellar mass.

We thus conclude that both minor merging and stellar mass loss are
responsible for the expansion of the dark matter halo since $z=2.1$
in our forced quenching simulations.

\section{Discussion}
\label{sec:discussion}

For the massive galaxies we simulate in this paper there are three
main processes that determine the halo response: dissipative (gas)
accretion; dissipationless (dry) merging and gas outflows driven by
AGN.  In lower mass galaxies SN/stellar winds are the dominant drivers
of gas outflows.  The first two processes are included in our
simulations, while AGN are not.  Dissipative accretion will only cause the
halo to contract, while dry merging and AGN feedback can (but will not
necessarily) cause halo expansion.  These processes are not
independent, as for example, stronger dissipation leads to more
contraction in the central galaxy, but also denser satellites which
can survive to smaller radii, and cause more expansion to the dark
matter. Thus, it is not obvious whether more dissipation will always
translate into more contraction.

\subsection{Does dry merging lead to halo expansion?}
In the halo quenching scenario the mass assembly is dominated since
$z\sim 2$ by dry mergers. It has been proposed (El-Zant \etal 2001;
Lackner \& Ostriker 2010) that this will create cores in the dark
matter halo. While we do find that dry merging causes haloes to expand
relative to the $z \sim 2$ case, the net effect at the halo mass scale
we study ($M_{200}\sim 10^{13}\Msun$) is still contraction: the dark
mass within 5 kpc has increased by factor of $\sim 1.4$.  Our
simulations show that the contractive effects of dissipation at early
times outweigh the expansive effects of dry merging at late times.  In
contrast to previous studies, our simulations are fully cosmological
with realistic progenitor masses and sizes which form realistic
elliptical galaxies at $z=0$. This is important, as the stellar
density, orbits, and number of satellite galaxies determines the
magnitude of the halo expansion effect.

\subsection{What is the halo response due to AGN feedback?}  Simulations that
incorporate AGN driven feedback can result in halo expansion (e.g.,
Peirani \etal 2008; Duffy \etal 2010; Martizzi \etal 2012), although
others seems to result in no significant change (e.g., Schaller \etal
2015a). Since our forced quenching simulations produce contracted
haloes, this difference in halo response suggests a new way to
distinguish between suppressive (halo) quenching and ejective (AGN)
quenching.

It still needs to be determined what simulations with AGN feedback
actually predict for halo structure. It is important that the
simulations reproduce the global properties of galaxies such as we
discuss in \S~\ref{sec:global} since the strength of the feedback will
effect both the halo response and the masses and structure of the
galaxies.  Theoretical models of AGN feedback are often calibrated to
match the present day galaxy stellar mass function (e.g., Schaye \etal
2015). A complication to this calibration is that the true stellar
masses of galaxies are not known accurately.  In the case of the
EAGLES simulations (Schaye \etal 2015) the observed stellar mass
function used in the calibration was derived assuming a  Milky Way
IMF. We know from dynamics and lensing studies that a Milky Way IMF
requires close to adiabatically contracted dark matter haloes in
elliptical galaxies (Auger \etal 2010a; Schulz \etal 2010; Dutton
\etal 2011b; 2013b). However, the halos in the EAGLES simulations show
almost no change (Schaller \etal 2015a), which points towards an
inconsistency between these simulations and observations. 

Furthermore, recent evidence points towards non-Milky Way IMFs in
massive elliptical galaxies from both dynamical and stellar population
modeling approaches (e.g., Auger \etal 2010a; van Dokkum \& Conroy
2010; Conroy \& van Dokkum 2012; Dutton \etal 2013a,b; Cappellari
\etal 2013; Barnab{\`e} \etal 2013; Ferreras \etal 2013; Sonnenfeld
\etal 2015).  The favored stellar masses are a factor of $\sim 2$
higher than obtained assuming a Milky Way IMF. Thus it would seem
necessary for simulations such as EAGLES to be re-calibrated to galaxy
mass functions based on heavier IMFs in the most massive galaxies. 

The structural properties of galaxies can provide additional
constraints for AGN feedback models. In particular the total mass
density slopes at $\sim 1\%$ of the virial radius, and the size vs
velocity relation are independent of uncertainties in the IMF.  At a
halo mass of $\sim 3\times 10^{13}\Msun$ the total mass density slopes
at 1\% of the virial radius in the EAGLES simulations are typically
shallower than observed (Schaller \etal 2015b), providing further
evidence for insufficient  dissipation.

Simultaneously reproducing several galaxy structural properties is a
non-trivial task.  Dubois \etal (2013) present six  zoom-in
simulations of halo mass $0.4 - 8 \times 10^{13}\Msun$ that include
AGN feedback. The simulations broadly reproduce the size vs stellar
mass, velocity dispersion vs stellar mass and total mass density
slopes at redshift $z\sim 0$. In detail there are some discrepancies
that again point towards insufficient dissipation. The total mass
density slopes are too shallow, and relative to observed stellar
masses derived using a Salpeter IMF the simulated sizes are too large
while the velocity dispersions are slightly too small.  

\subsection{Future directions for halo quenching models}
The halo quenching model we implement in this paper could be improved
upon in several ways. We turn the gas adiabatic everywhere in the
simulation at $z\sim 2$, which corresponds to when the most massive
progenitor reaches a halo mass of $\sim 10^{12}\Msun$.  A more
realistic scenario would be to only shut off cooling in haloes above
$\sim 10^{12}\Msun$, or some other galaxy property such as stellar
density or bulge mass, allowing cooling and star forming in lower mass
progenitor galaxies to occur below $z=2.1$.  This additional cooling
would presumably result in denser inner galaxies, and more
contraction. Another improvement would be to implement
photo-ionization from stellar sources (Kannan \etal 2014, 2015) or
heating from AGB stars (Conroy \etal 2015) directly into the
simulations.

Simulations could be run at different halo masses, e.g., from $\sim
10^{12} - 10^{14}\Msun$ to determine if there is any mass dependence
to the halo contraction in halo quenching scenario. In the context of
$\LCDM$, more massive haloes have a larger fraction of the stellar
mass in the central galaxy assembled through mergers (Behroozi \etal
2013), and thus the expansive effects of dissipationless assembly
would be expected to be more important in higher mass haloes.
Indeed, at the galaxy cluster scale $M_{200}\sim 10^{15}\Msun$ Laporte
\& White (2015) find that dissipationless assembly since $z=2$
reverses the contraction in the progenitor galaxies resulting in  net
halo expansion within $\sim 1\%$ of the virial radius by redshift
$z=0$. We note that at a halo mass of $10^{13}\Msun$ the accreted
fraction of stars is already quite high at $\sim 60\%$ (Behroozi \etal
2013), thus any mass dependence is likely to be weak. Furthermore,
since the halo and galaxy mass functions are steep, very few quiescent
galaxies live in the highest mass dark matter haloes, and thus
majority of massive elliptical galaxies will have contracted dark
matter haloes in the halo quenching scenario.

\subsection{Observations of dark matter fractions}
An observational test of halo response models is the dark matter
fraction on scales of galaxy half-light radii, i.e.,  $\sim 5$
kpc. Total masses on these scales can be measured reliably using
stellar kinematics and/or strong gravitational lensing (e.g., see
Courteau \etal 2014 for a review). However, decomposing the total mass
into baryons and dark matter is more challenging. Stellar population
synthesis models can be used to convert luminosity profiles into
stellar mass profiles up to the stellar IMF and systematics uncertain
phases in stellar evolution.  It is well established observationally
that if the IMF is universal, then massive elliptical galaxies have
dark matter fractions of $f_{\rm DM}(R_{e})\sim 0.5$ and in the
context of $\LCDM$, require close to adiabatic halo contraction (Auger
\etal 2010a; Schulz \etal 2010; Dutton \etal 2011b, 2013b).

\begin{figure}
\centerline{
\psfig{figure=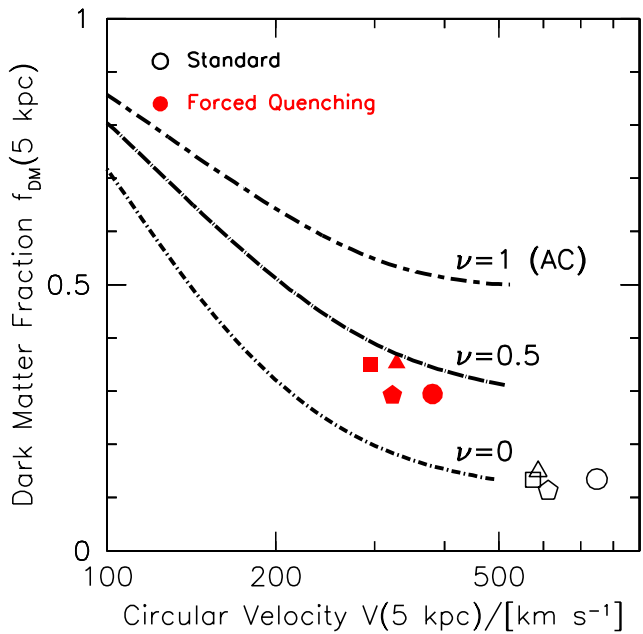,width=0.5\textwidth}
}
\caption{Dark matter fraction vs circular velocity, both measured at 5
  kpc. Symbols show our simulations while lines show observed values
  for different halo response models from Dutton \etal (2013b): no
  contraction ($\nu=0$), weak contraction ($\nu=0.5$)  adiabatic
  contraction (AC, $\nu=1$). The forced quenching simulations (red)
  are most similar to the weak contraction model. The standard
  simulations (black) have very low dark matter fractions, which
  observationally requires uncontracted halos, but the simulations
  have contracted indicating an inconsistency.}
\label{fig:fdm}
\end{figure}

In Fig.~\ref{fig:fdm} we show the dark matter fractions within spheres
of radius 5 kpc vs the circular velocity at 5 kpc in our simulations
(symbols) and compared to observationally constrained models (lines)
from Dutton \etal (2013b). This plot is similar to Fig.~21 of Dubois
\etal (2013), but we use a fixed physical radius, rather than a
relative radius such as the half-light radius to eliminate any
differences of scale between the simulations and observations. The
forced quenching simulations have dark matter fractions of $\sim
30\%$, while the standard simulations have $\sim 10\%$.  The lines
show models with different assumptions about the halo response
parameterized by Eq.~\ref{eq:acnu} from no change ($\nu=0$) to
adiabatic contraction ($\nu=1$). Models with stronger contraction have
correspondingly lower stellar mass to light ratios. At a circular
velocity of $300 \kms$ the stellar mass offsets are $0.23, 0.16$, and
$0.03$ dex, where 0.23 corresponds to a Salpeter IMF, and 0.0 to a
Chabrier IMF.  Note that all models reproduce (by construction) the
scaling relations between size, stellar mass, and velocity
dispersion. 

An accurate measurement of the dark matter within 5 kpc would thus be
able to falsify our simulations for elliptical galaxy formation.  We
note that variety of studies seem to favor models closer to $\nu=0$
and Salpeter IMFs than $\nu=1$ and Chabrier IMFs (Auger \etal 2010a;
Dutton \etal 2013b; Dutton \& Treu 2014; Sonnenfeld \etal 2015),
although there are some exceptions (Smith \etal 2015). Clearly more
work needs to be done to make the distinction more definitive.

\subsection{Priors on dark matter density slopes}
Our results have implications for mass modeling of elliptical
galaxies.  Observations commonly parameterize the dark matter halo as
either a power-law with a free slope, or
a double power-law with a free inner slope:
\begin{equation}
\label{eq:power2}
  \rho(r)/\rho_0 = (r/r_{\rm s})^{-\alpha}(1 + r/r_{\rm s})^{-3+\alpha}.
\end{equation}
For example, Cappellari \etal (2013) adopt Eq.~\ref{eq:power2} with a
uniform prior on inner slope with limits [0,1.6].  The upper limit is
consistent with our forced quenching simulations, but excludes
simulations with stronger contraction.  Due to the degeneracy between
the halo response and dark matter fraction  with the stellar
mass-to-light ratio (e.g., Dutton \etal 2013b) the choice of prior is
important if one wishes to accurately constrain stellar mass-to-light
ratios and make inferences on the IMF.  Our simulations suggest an
upper limit of $\alpha=1.6$ is too restrictive.  A more conservative
upper limit of $\alpha=2$ allows for the full range of inner slopes
found in our simulations.

\section{Summary}
\label{sec:summary}

We use cosmological hydrodynamical simulations with the SPH code {\sc
  gasoline} to investigate the response of dark matter haloes to the
formation of massive elliptical galaxies.  We consider 4 initial
conditions that grow into haloes of mass $M_{200}\sim 10^{13}\Msun$ by
the present day.  At our standard resolution each simulation has more
than 1.5 million dark matter particles within the virial radius at
$z=0$ enabling us to resolve the dynamics at 1\% of the virial radius.
Our highest resolution simulation has $10.6$ million dark matter
particles within the virial radius at $z=0$, making it one of the
highest resolution simulations of elliptical galaxy formation to date.
Our standard simulations include metallicity dependent gas cooling,
star formation, and stellar feedback.

We summarize our results as follows:
\begin{itemize}

\item  At redshift $z=2.1$ our standard simulations have stellar to
  halo mass ratios consistent with halo abundance matching, assuming a
  Salpeter IMF to derive the stellar masses (Fig.~\ref{fig:mm}).
  Galaxy half-mass sizes and circular velocities are also consistent
  with observations (Fig.~\ref{fig:rmvm}).

\item At $z=2.1$ the dark matter haloes have contracted in response to
  galaxy formation
  (Figs.~\ref{fig:vcirc_fid_z2},~\ref{fig:vratio},~\ref{fig:ac}), but
  not as strong as predicted by the adiabatic contraction formalism
  (Blumenthal \etal 1986).

\item By $z=0$ the standard simulations have overcooled, with a factor
  of $\gta 4$ times too many stars, resulting in circular velocities,
  stellar densities, and mass density slopes that are too high
  (Figs.~\ref{fig:mm}$-$\ref{fig:vv})

\item We investigate the halo quenching scenario by shutting down
  cooling and star formation at $z=2.1$ (when the most massive
  progenitors have $M_{200}\sim 10^{12}\Msun$) and evolving the
  simulation hierarchically to $z=0$. We refer to these simulations as
  forced quenching (FQ).  The resulting galaxies have many properties
  consistent with observed elliptical galaxies: $\Mstar\sim 2\times
  10^{11}\Msun$, $\Mstar/M_{200}\sim 0.01$ (Fig.~\ref{fig:mm}), flat
  circular velocity profiles  with mass density slope $\gamma'\sim 2$
  (Figs.~\ref{fig:gammap} \& \ref{fig:vcirc_hq_z0}), half-mass sizes
  of $r_{1/2}\sim 4-10$ kpc (Fig.~\ref{fig:rmvm}), and circular
  velocities $V_{\rm circ}(r_{1/2})\sim 300 -400 \kms$
  (Fig.~\ref{fig:rmvm}).

\item In all the FQ simulations the dark matter haloes contract,
  although less than in the standard simulations
  (Fig.~\ref{fig:vratio}). The contraction is much weaker than
  predicted by the adiabatic contraction models of Blumenthal \etal
  (1986) and Gnedin \etal (2004),  but can be described with a simple
  formula (Eq.~\ref{eq:fq0}).

\item The dark matter density slopes (at redshift $z=0$) measured
  between 1-2\% of the virial radius vary from $-1.2\pm0.1$
  in the control simulations to $-1.53\pm0.17$ in the FQ simulations and
  to $-1.76\pm 0.17$ in the standard simulations
  (Fig.~\ref{fig:dicintio}).
  
\item  Dry merging alone is unable to reverse the contractive effects
  of early dissipation, which must occur to form an  elliptical
  galaxy.
  
\item Simulations in the literature find that AGN feedback can cause
  halo expansion (e.g., Duffy \etal 2010; Martizzi \etal 2012), and
  thus there may be qualitatively different halo responses, and dark
  matter fractions within $\sim 5$ kpc, in the suppressive (halo)
  quenching and ejective (AGN) quenching scenarios. We note that
  different halo responses will also require different stellar IMFs
  in order to be consistent with observed constraints on dark matter
  fractions (Fig.~\ref{fig:fdm}).
  
\end{itemize}

While our treatment of the halo quenching process is clearly an
oversimplification, we hope that the excellent agreement between the
structural properties of our forced quenching simulations with
observations motivates further studies of the halo quenching scenario.

\section*{Acknowledgments} 
We thank the referee for a prompt and constructive report.  The
simulations were performed on the theo cluster of the
Max-Planck-Institut f\"ur Astronomie at the Rechenzentrum in Garching.
We acknowledge support from the Sonderforschungsbereich SFB 881 ``The
Milky Way System'' (subproject A01 for AAD, AVM, GSS, and TG, and
subproject A02 for CP and TB) of the German Research Foundation (DFG).


\appendix
\section{galaxy parameters}
Table~\ref{tab:data} lists the parameters of the most massive galaxy in each simulation at redshift $z\sim 2$ and $z\sim 0$.
Note that the actual redshifts vary slightly to avoid major mergers that occur at the nominal output redshifts.

\begin{table*}
 \centering
 \caption{Galaxy parameters. Column (1) name of the initial conditions
   -- see Table~\ref{tab:setup} for simulation parameters. Column (2)
   type of simulation: standard (S) or forced quenching (FQ). Column
   (3), $z$, redshift of output. Column (4), $M_{200}$,   virial
   mass. Column (5), $\Mstar$, stellar mass inside 20\% of the virial
   radius. Column (6), $R_{1/2}$, 3D half-mass radius. Column (7),
   $V_{1/2}$, circular velocity ($V(r)=\sqrt{GM(r)/r}$) at the 3D
   half-mass radius.  Column (8), $V_{200}$, circular velocity at the
   virial radius. Column (9), $V_{5\rm kpc}$, circular velocity at 5
   kpc.  Column (10), $f_{\rm DM}$, dark matter fraction at 5
   kpc. Column (11), $\alpha$, slope of the dark matter density
   profile  measured between 1 and 2\% of the virial radius.  Column
   (12), $\gamma'$, slope of the total mass profile measured between 1
   and 2\% of the virial radius.  Column (13), $\Sigma_{\rm star}$,
   average surface density of the stars inside the 2D half-stellar
   mass radius.}
  \begin{tabular}{lcccccccccccc}
\hline
\hline  
Name & type & $z$ & $\log_{10}M_{200}$ & $\log_{10}M_{\rm star}$ & $R_{1/2}$ & $V_{1/2}$ & $V_{200}$ & $V_{5 \rm kpc}$ & $f_{\rm DM}$ & $\alpha$ & $\gamma'$ & $\log_{10}\Sigma_{\rm star}$\\ 
     & & & [M$_{\odot}$] & [M$_{\odot}$] & [kpc] & [$\kms$] & [$\kms$] & [$\kms$] &  &  &  & $[\Msun {\rm pc}^{-2}]$ \\
(1) & (2) & (3) & (4) & (5) & (6) & (7) & (8) & (9) & (10) & (11) & (12) & (13)\\
\hline
halo1   & S & 0.12 & 13.14 & 11.80 & 4.28 & 599.3 & 355.5 & 589.5 & 0.146 & -1.88 & 2.37 & 3.99\\
halo2   & S & 0.10 & 13.14 & 11.82 & 4.68 & 580.7 & 353.7 & 577.4 & 0.130 & -1.93 & 2.42 & 3.93\\
halo3   & S & 0.08 & 13.14 & 11.94 & 5.74 & 602.7 & 353.5 & 613.6 & 0.110 & -1.58 & 2.28 & 3.88\\
halo4.0 & S & 0.00 & 13.37 & 12.09 & 4.21 & 825.1 & 414.2 & 798.6 & 0.094 & -1.55 & 2.42 & 4.29\\
halo4.1 & S & 0.12 & 13.33 & 12.09 & 5.08 & 756.7 & 409.6 & 758.6 & 0.108 & -1.61 & 2.36 & 4.13\\
halo4.2 & S & 0.00 & 13.34 & 12.07 & 5.13 & 744.4 & 403.9 & 747.6 & 0.135 & -1.67 & 2.36 & 4.10\\
\hline
halo1   &FQ & 0.00 & 13.14 & 11.32 &  6.45 & 343.3 & 348.4 & 326.0 & 0.349 & -1.71 & 1.80 & 3.17\\
halo2   &FQ & 0.14 & 13.13 & 11.00 &  3.59 & 288.2 & 354.2 & 295.3 & 0.344 & -1.63 & 2.10 & 3.36\\
halo3   &FQ & 0.00 & 13.19 & 11.41 &  7.95 & 340.5 & 362.3 & 321.1 & 0.288 & -1.36 & 1.79 & 3.09\\
halo4.0 &FQ & 0.00 & 13.36 & 11.09 & 13.79 & 331.5 & 410.9 & 253.4 & 0.564 & -1.51 & 1.49 & 2.26\\
halo4.1 &FQ & 0.00 & 13.35 & 11.42 &  7.01 & 369.5 & 408.0 & 381.6 & 0.310 & -1.43 & 2.08 & 3.18\\
halo4.2 &FQ & 0.00 & 13.33 & 11.44 &  7.78 & 372.1 & 401.2 & 380.6 & 0.295 & -1.52 & 2.03 & 3.11\\
halo4.3 &FQ & 0.00 & 13.43 & 11.48 & 11.08 & 376.9 & 432.3 & 348.3 & 0.363 & -1.57 & 1.82 & 2.84\\
\hline
halo1   & S & 2.22 & 12.15 & 11.01 &  2.06 & 384.7 & 237.0 & 398.0 & 0.390 & -1.20 & 1.21 & 3.83\\
halo2   & S & 2.08 & 12.19 & 10.92 &  1.43 & 404.6 & 240.6 & 373.8 & 0.360 & -1.61 & 1.58 & 4.06\\
halo3   & S & 2.08 & 12.32 & 10.94 &  1.51 & 385.1 & 265.3 & 363.8 & 0.356 & -1.57 & 1.63 & 4.03\\
halo4.0 & S & 2.07 & 12.30 & 10.79 &  2.69 & 320.4 & 259.9 & 343.3 & 0.404 & -1.98 & 0.99 & 3.38\\
halo4.1 & S & 2.07 & 12.25 & 11.14 &  0.96 & 582.2 & 251.2 & 435.0 & 0.354 & -1.86 & 2.31 & 4.62\\
halo4.2 & S & 2.22 & 12.22 & 11.13 &  1.30 & 517.2 & 251.6 & 437.1 & 0.383 & -1.76 & 1.93 & 4.35\\
halo4.3 & S & 2.07 & 12.42 & 11.04 &  1.94 & 410.0 & 286.8 & 401.0 & 0.402 & -1.59 & 1.79 & 3.92\\
\hline
\hline
\label{tab:data}
\end{tabular}
\end{table*}

\label{lastpage}
\end{document}